\documentclass[aps,pra,reprint,superscriptaddress]{revtex4-2}
\usepackage{graphicx}
\usepackage{color}
\usepackage{amsmath,amssymb}
\usepackage{bm}
\usepackage{upgreek}
\usepackage{xspace}
\usepackage[colorlinks,urlcolor=blue,citecolor=blue,linkcolor=blue]{hyperref}

\usepackage{bbold}

\usepackage[english]{babel}

\begin{document}

\title{Wavelength dependence of nitrogen-vacancy center charge cycling}

\author{A.~A.~Wood}
\email{alexander.wood@unimelb.edu.au}
\affiliation{School of Physics, University of Melbourne, Victoria 3010, Australia}
\author{A. Lozovoi}
\affiliation{CUNY-The City College of New York, New York 10031, USA}
\author{R. M. Goldblatt}
\affiliation{School of Physics, University of Melbourne, Victoria 3010, Australia}
\author{C. A. Meriles}
\affiliation{CUNY-The City College of New York, New York 10031, USA}
\author{A. M. Martin}
\affiliation{School of Physics, University of Melbourne, Victoria 3010, Australia}

\date{\today}
\begin{abstract}
Optically-active spin qubits in wide-bandgap semiconductors exist in several charge states, though typically only specific charge states exhibit desirable spin or photonic properties. An understanding of how interconversion between different charge states occurs is important for most applications seeking to employ such defects in quantum sensing and information processing, and additionally serves as a means of testing and verifying models of the defect electronic structure. Here, we use charge-sensitive confocal imaging to study the wavelength dependence of optical carrier generation in diamonds hosting nitrogen-vacancy (NV) centers, silicon vacancy (SiV) centers and substitutional nitrogen (N). We study the generation of distinctive charge-capture patterns formed when photogenerated charge carriers are captured by photoluminescent defects, using light spanning 405-633\,nm (1.96-3.06\,eV). We observe distinct regimes where one- or two-photon ionization or recombination processes dominate, and a third regime where anti-Stokes mediated recombination drives weak NV charge cycling with red light. We then compare red-induced charge cycling to fast charge carrier transport between isolated single NV centers driven with green and blue light. This work reports new optically-mediated charge cycling processes of the NV centers, and has consequences for schemes using charge transfer to identify non-luminescent defects and photoelectric detection, where ambiguity exists as to the source of photocurrent. 
\end{abstract}
\maketitle
\section{Introduction}
While many optically-active and spin-addressable defects exist in materials such as diamond~\cite{doherty_nitrogen-vacancy_2013, thiering_chapter_2020}, silicon carbide~\cite{castelletto_silicon_2020} and hexagonal boron nitride~\cite{tran_robust_2016, caldwell_photonics_2019}, the spin and photonic properties that are sought are often only associated with a particular charge state of the defect, which may in turn exhibit instability under optical addressing~\cite{aslam_photo-induced_2013}. For instance, the neutral charge state of the nitrogen-vacancy (NV) center in diamond (NV$^0$) with a zero-phonon line (ZPL) at 575\,nm exhibits neither the spin-dependent fluorescence or long ground-state coherence times of the negatively-charged NV$^-$ (ZPL at 637\,nm) that features prominently as a platform for quantum sensing and information processing~\cite{waldherr_dark_2011}. Besides the NV center, many other defects exist with various charge states that exhibit photochromism, like the silicon-vacancy SiV$^-$, SiV$^0$ and SiV$^{2-}$ charge states~\cite{rose_observation_2018, wood_room-temperature_2023, zhang_neutral_2023}. On the other hand, other defects that are optically inactive, such as substitutional nitrogen which exists in N$^0$, N$^+$ and N$^-$ states~\cite{ashfold_nitrogen_2020} have important consequences for the charge stability and spin coherence of nearby color centers~\cite{manson_NV-_2018, lozovoi_dark_2020, wang_manipulating_2023} due to their abundance. 

Charge state measurement and control represents a powerful tool to augment and improve existing quantum measurement techniques~\cite{barry_sensitivity_2020} as well as facilitating new applications~\cite{han_metastable_2010,shields_efficient_2015, dhomkar_long-term_2016, pfender_protecting_2017, mccloskey_diamond_2022, monge_reversible_2023, ji_correlated_2024, delord_correlated_2024}, in particular, photoelectric detection of NV magnetic resonance~\cite{bourgeois_photoelectric_2015, siyushev_photoelectrical_2019, bourgeois_photoelectric_2020}. It is in photoelectric detection that knowledge of the charge interconversion processes are most important, such as the role of other defects that contribute to the photocurrent~\cite{hruby_magnetic_2022, bourgeois_photoelectric_2022}, particularly under varied excitation wavelengths~\cite{todenhagen_wavelength_2023}. Similarly, the role of freely diffusing charges generated by optical excitation has been examined in observation and control of the space-charge potentials accompanying the generation and diffusion of free carriers~\cite{lozovoi_probing_2020}, identification of charge carriers from optical illumination of different defect species~\cite{gardill_probing_2021, lozovoi_imaging_2022}, transport of charge carriers between single NV defects~\cite{lozovoi_optical_2021, lozovoi_imaging_2022, lozovoi_detection_2023} and control over the charge environment-limited spin coherence of NV centers~\cite{lozovoi_dark_2020, zheng_coherence_2022, wang_manipulating_2023}

In this work, we investigate the photogeneration and capture of charge carriers in three diamond samples containing varying densities of nitrogen, NV centers, and SiV centers as a function of excitation wavelength. We study the formation of characteristic `halo' patterns of carriers diffusing into charge environments tailored to specific defect concentrations via multi-wavelength scanning confocal microscopy. The range of excitation wavelengths we use span 405-633\,nm, sufficient to observe single-photon mediated photoionization and recombination of NV centers in the blue (405-480\,nm), two-photon charge cycling of the NV charge state (515-594\,nm), and weak two-photon recombination of NV$^0$ and near-resonant ionization of NV$^-$ (594, 633\,nm), which we ascribe to phonon-assisted anti-Stokes excitation of the NV$^0$ zero-phonon line (ZPL) transition. We do not find any evidence that SiV defects alone undergo charge cycling under optical illumination, even under red excitation, with silicon-free diamonds exhibiting essentially the same charge carrier generation as SiV-rich samples. We then show that red excitation of a single NV center mediates hole generation and capture by a second nearby single NV center, and compare this with carrier generation under blue and green light, observing carrier generation and transport rates an order of magnitude faster than previously reported~\cite{lozovoi_optical_2021}. Our work has important consequences for the study of charge carrier generation and capture in diamond, which spans photoelectric detection~\cite{bourgeois_photoelectric_2020, bourgeois_photoelectric_2022}, stabilization of defect charge states~\cite{gorlitz_coherence_2022, zuber_shallow_2023} and the identification of photoactive and inactive defects~\cite{gardill_probing_2021, lozovoi_imaging_2022}.

\section{Experiment}
\begin{figure}
	\centering
		\includegraphics[width = \columnwidth]{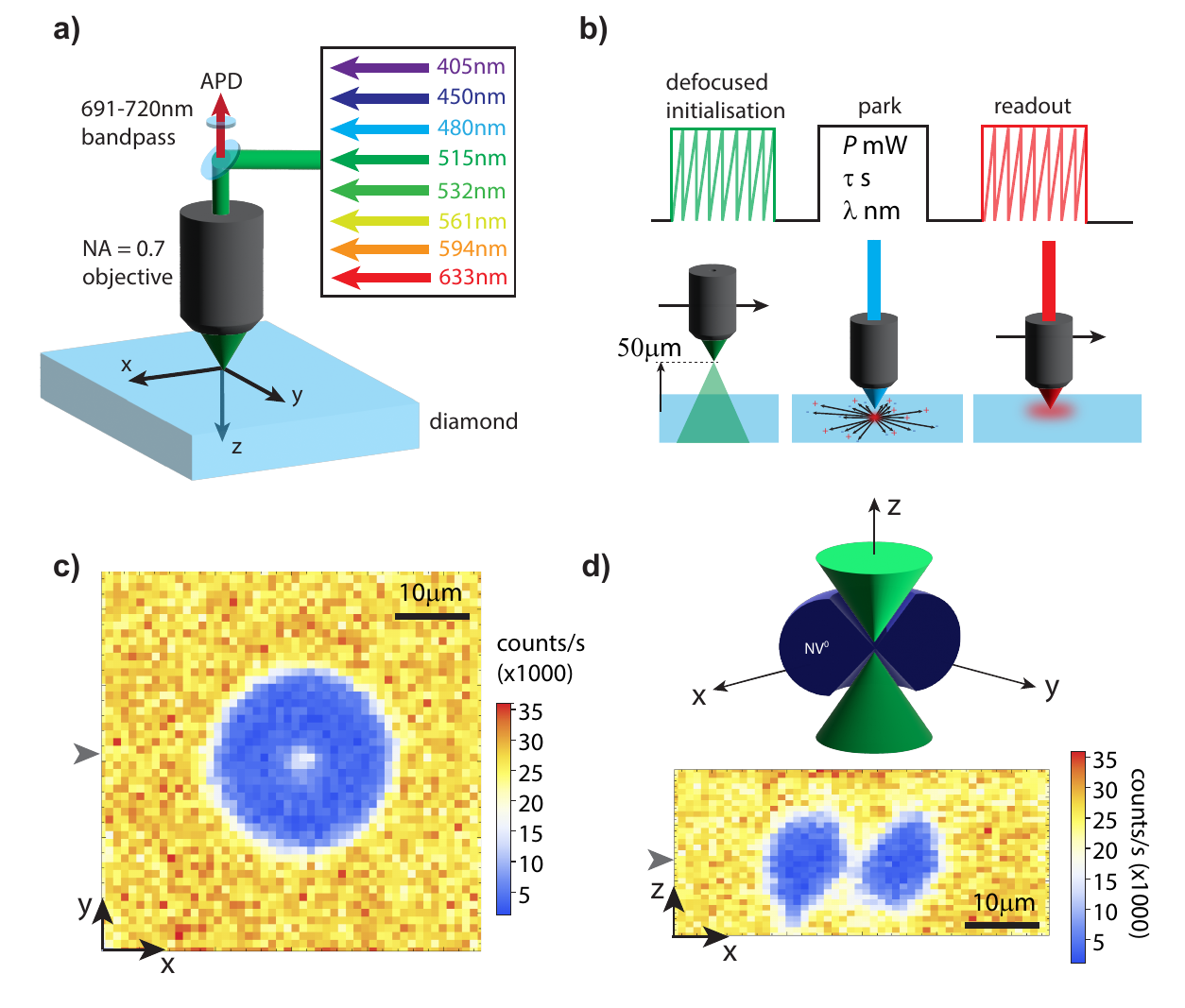}
	\caption{Photogeneration of charge carriers and measurement of charge states. a) Schematic of the setup, showing microscope objective on three-axis scanning stage, 6\,mm working distance positioned above the diamond and 8 illumination wavelengths spanning 405-633\,nm. b) Protocol for charge initialization of NV$^-$ charge state, showing defocused confocal scanning 50\,$\upmu$m above the imaging plane (10\,$\upmu$m below the diamond surface), followed by parking a laser with wavelength $\lambda$, power $P$ for time $\tau$ in the imaging plane. The resulting distribution of charge states is read out with a weak ($50\,\upmu$W) red scan. c) $x$-$y$ and d) $x$-$z$ scans of NV$^-$ fluorescence following a $P=$8\,mW, $\tau = 100\,$ms $\lambda = 532$\,nm park. Grey arrows denote the $z$ and $y$ positions in the $x$-$y$ and $x$-$z$ scans. Top panel of d) shows a 3D schematic of the spherical NV$^0$ charge distribution, with the focused green laser locally generating NV$^-$ above and below the imaging plane.}
	\label{fig:fig1}
\end{figure}

A simplified schematic of our experiment is depicted in Fig \ref{fig:fig1}(a). We use a home-built scanning confocal microscope that features optical paths for excitation wavelengths ranging from 405\,nm to 633\,nm, which are directed into the back aperture of a 0.7\,NA microscope objective mounted on a three-axis scanning piezoelectric stage. The different colors of light are sourced from a variety of fixed-wavelength diode or diode-pumped solid state lasers that are intensity controlled by either acousto-optic modulators (AOMs: 480\,nm, 532\,nm, 561\,nm and 594\,nm) or by directly varying the diode drive current (405\,nm, 450\,nm, 515\,nm and 633\,nm). Fluorescence from the diamond is collected by the same objective lens, directed through bandpass filters to isolate NV$^-$ emission (691-730\,nm) and then focused by a 150\,mm-focal-length achromatic lens into a $50\,\upmu$m-core-diameter multimode fiber, where it is directed into a single-photon counting module.

The diamond samples we investigated were obtained from commercial suppliers, grown via chemical vapor deposition (CVD), and contain a natural abundance of $^{13}$C. Sample A is a [100]-cut electronic-grade commercial diamond from Element6 with [N] = $<0.1$ppm, [NV] = 10 ppb and no detectable SiV fluorescence. Sample B is a [111]-cut standard-grade sample from Delaware Diamond Knives with [N] = 1\,ppm, [NV] = 10\,ppb and [SiV] = 1\,ppm (estimated). Aside from crystallographic orientation and SiV concentrations, the samples feature similar fluorescence levels under green illumination and similar spin coherence properties $(T_2^\ast = 0.5-1\,\upmu$s, $T_2 = 200-300\,\upmu$s), and feature nitrogen as the dominant impurity~\cite{edmonds_production_2012}.

Our experimental procedure, shown in Fig. \ref{fig:fig1}(b) follows three steps: \emph{initialization} of defects into a particular charge state by scanning the optical illumination across a region of the diamond, followed by \emph{parking} the illumination point at a fixed position with excitation wavelength $\lambda$, power $P$ for duration $T$, and then finally a \emph{readout} scanning step with weak red light (633\,nm, 50\,$\upmu$W) to read out the spatial charge distribution. 

\emph{Charge initialization. }\label{sec:ssecinit}
Scanning green light over a given spatial region at set intensities initializes that region into a given defect charge state concentration. Immediately under the laser spot, green illumination favors the production of NV$^-$. However at higher green powers, rapid generation of holes under the illumination spot from NV charge cycling~\cite{dhomkar_-demand_2018} creates a halo-like region surrounding the laser spot of NV$^0$ due to hole capture, as shown in Fig. \ref{fig:fig1}(c). As the laser is scanned in position, it leaves a predominantly NV$^0$ distribution in its wake. Thus, strong green scans prepare an NV$^0$-rich, optically-dark region.

By elevating the microscope objective some $40\,\upmu$m above the diamond surface during green illumination scans we create a large, low intensity illumination spot with negligible charge carrier generation compared to a tightly focused beam in the imaging plane. This `defocused preparation' step is observed to maximize NV$^-$ generation, which we attribute to sufficiently low intensities to photoionize N$^0$ ($\propto I$) without instigating charge cycling of the NVs ($\propto I^2$). With continual photoionization of N$^0$, free electrons are trapped by NV$^0$ after sufficiently long illumination times ($>$1-2\,s). Scanning the laser beam broadens the effective preparation region, resulting in a bright, NV$^-$ rich preparation. 

For sample B, containing a significant SiV fraction, strong green scans generate SiV$^-$~\cite{wood_room-temperature_2023}, while the defocused preparation step is observed to favor production of a NV$^-$ and SiV$^{2-}$ rich charge distribution, possibly via similar association of extra electrons to SiV$^-$. The formation of SiV$^{2-}$ can also be attributed to the green photon energy (2.33\,eV) exceeding the ionization threshold of SiV$^-$ (2.1\,eV)~\cite{gali_ab_2013, thiering_ab_2018}, and since NV charge cycling is minimized during defocused scanning, no process exists to generate holes to recombine to form SiV$^-$. Similarly, strong green scanning in the imaging plane generates holes that deplete NV$^-$ and produce SiV$^-$. 

Scanning the laser beam position allows us to prepare regions of up to $(200\,\upmu\text{m})^2$ in a given charge state distribution. A single initialization step consists of a strong (8\,mW) green scan followed by a defocused preparation, at 8\,mW but with the objective $50\,\upmu$m above the diamond surface.

\emph{Parking. }
Following initialization, we then apply a single laser pulse at a fixed location to induce local charge state conversion and the generation of free carriers. The parking step typically generates a dark charge-carrier capture pattern (CCP), or `halo', around the illumination region as fluorescent defects (NV$^-$) capture holes and transition to non-fluorescent charge states. Fig. \ref{fig:fig1}(c) shows the salient features of these patterns formed under green illumination and imaged on a bright NV$^-$ background. Later in Section \ref{sec:wavelength}, we will principally focus on the radius of this halo feature to quantitatively assess the rate of charge photogeneration. In previous work, charge carrier generation under green (532\,nm and 514\,nm) and red (633\,nm) has been studied~\cite{jayakumar_optical_2016, jayakumar_long-term_2020, gardill_probing_2021, lozovoi_optical_2021}, with the latter excitation wavelength suspected to induce charge cycling of either an unknown defect~\cite{lozovoi_imaging_2022} or of SiV centers~\cite{zhang_neutral_2023}. In this work, we study charge generation and capture using several key additional illumination wavelengths for the first time. Blue light (405-480\,nm) induces single-photon mediated ionization and recombination for both NV$^-$ and NV$^0$~\cite{bourgeois_enhanced_2017} and subsequent charge cycling. Light at 561\,nm is believed to excite NV charge cycling though without photoionization of nitrogen impurities~\cite{hruby_magnetic_2022}, and 594\,nm orange light is below the zero-phonon line excitation threshold for NV$^0$ (575\,nm), and as such should impair photorecombination of NV$^0$ back to NV$^-$, significantly altering NV charge cycling.
       
\emph{Readout. }    
In all experiments described in this work, we read out the charge distribution pattern arising from a laser parking step with a weak red scan, typically $50\,\upmu$W. At these powers, the intensity of the red light is low enough to prevent significant ionization of NV centers (NV$^-\rightarrow\text{NV}^0$) though still high enough to excite fluorescence. We limit the collected fluorescence to only NV$^-$ (691-730\,nm), excluding SiV$^-$ emission (738\,nm) and resulting in very high contrast between NV$^-$ (0.5-1$\times10^5$ counts/s) and NV$^0$ ($<5\times10^3$ counts/s). Parking the green laser on Sample B generates a bright SiV$^-$ CCP with an inner dark region due to SiV$^0$~\cite{wood_room-temperature_2023, zhang_neutral_2023}. However, for direct comparison with Sample A, which exhibits no SiV fluorescence, we study only the NV charge populations.  
                
The three-axis scanning capability of our microscope enables visualization of the depth ($z$) dependence of the CCP. In Figure \ref{fig:fig1}(d), we present an $x$-$z$ scan of the pattern, which appears to show lobes either side of the illumination region. This apparent asymmetry between $x$-$y$ and $x$-$z$ profiles of charge carrier photogeneration is a consequence of the axial expansion of the green laser exciting the NV centers. At the beam waist, the intensity is sufficient to excite photogeneration. However, the rapid expansion of the beam along the $z$-axis results in a green intensity sufficient to only optically prepare NV$^-$ a few microns above the waist, resulting in a `punctured-sphere' fluorescence distribution as depicted in the upper panel of Fig. \ref{fig:fig1}(d). 

\section{Wavelength dependence of photogeneration}\label{sec:wavelength}
\begin{figure*}[t!]
	\centering
		\includegraphics[width = \textwidth]{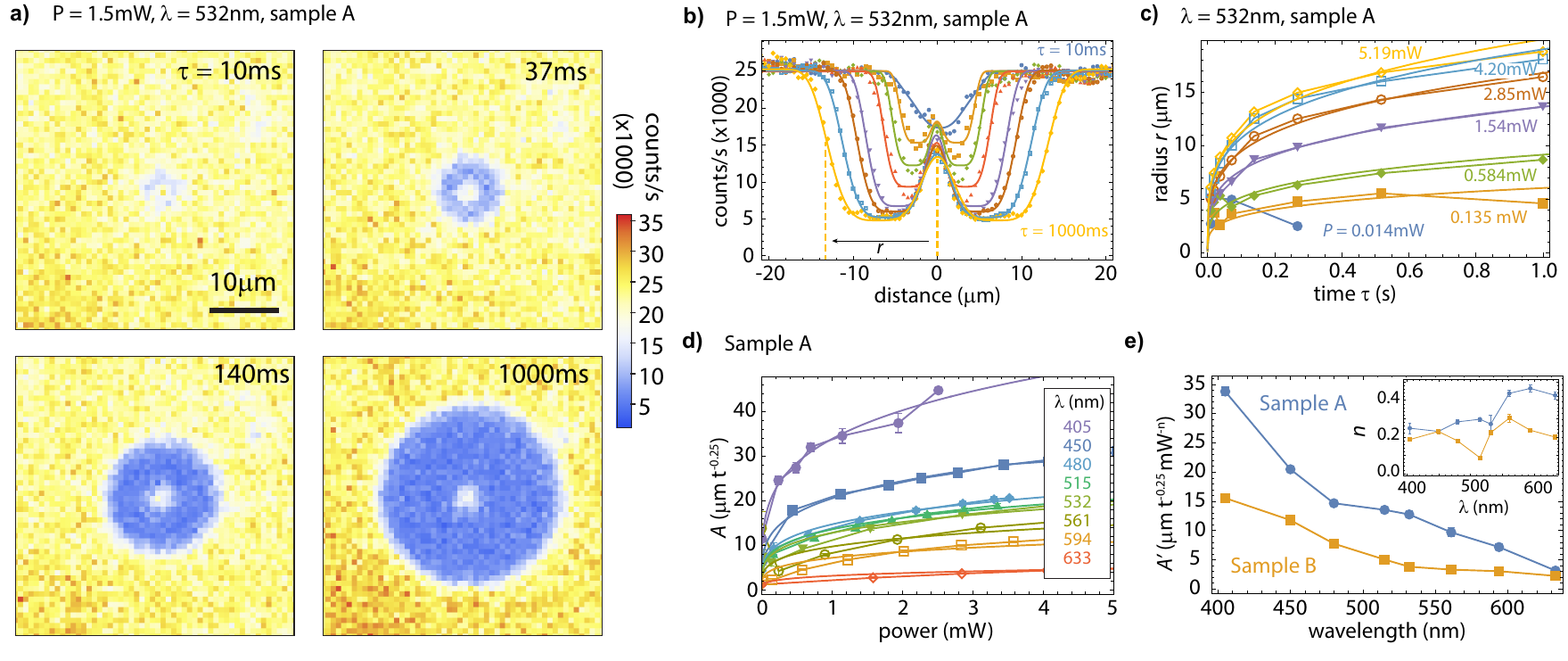}
	\caption{Charge generation as a function of time, power and wavelength. a) Exemplar confocal images of sample A NV$^-$ PL for a $P=$1.5\,mW, $\lambda = $532\,nm park for varied times. b) Averaged lineouts (data points) and double-quartic-gaussian fits (solid lines) for the complete dataset in (a), for $\tau =10\,$ms to $\tau = 1000\,$ms. (c) Fitted halo radii $r$ as a function of time for increasing 532\,nm park power $P$ (data) and fits of the form $r(\tau) = A t^{1/4}$ fits (solid lines). (d) Repeating the analysis across all excitation wavelengths, we extract $A$ as a function of power, plotted here for each park laser wavelength $\lambda$. Solid lines are fits of the form $A' P^n$, and in (d) we plot $A'$ and $n$ as a function of $\lambda$ for both samples. Halos grow faster and larger for lower excitation wavelengths, and denser samples admit smaller halos.}
	\label{fig:fig2}
\end{figure*}

With our experimental procedure now fully described, we turn to an examination of the wavelength dependence of the charge carrier photogeneration process responsible for NV$^0$ generation. In lieu of a high-power tunable laser source spanning the blue to near-IR, we select here several key illumination wavelengths easily sourced from commercial laser diodes or diode-pumped solid state systems. To parametrize the rate of charge carrier photogeneration, we consider as our figure of merit the radius $r$ of the NV$^0$ region, and study how this radius increases as a function of time, power and wavelength. Figure \ref{fig:fig2}(a) shows example confocal scans of NV$^0$ generation following a $P = 1.5$\,mW, $\lambda = 532$\,nm park for varied time. We consider an averaged 1D-lineout of the pattern and fit a function of the form $g(x) = A\left[ \exp(-(x-d)^4/\sigma^4)+\exp(-(x+d)^4/r^4)\right] + A_0$ to extract $r = \sigma + d$, the characteristic radius, as shown in Fig. \ref{fig:fig2}(b). The double-quartic-gaussian function is chosen empirically due to the flat-bottomed profile of the halo patterns and the bright central maximum, which originates from NVs directly under the illumination spot. We then plot $r$ vs. $t$ for a selection of powers in Fig. \ref{fig:fig2}(c).

The expansion of the radius $r$ as a function of time for all powers is observed to follow a $t^{1/4}$-like dependence. A fit to the data of the form $r(t) = A t^n$ with free $n$ reveals little to no significant trend or variation in $n$ across the entire dataset of powers, wavelengths and both samples, consistent with the same underlying processes generating the halo pattern (i.e. charge generation, diffusion and capture but at different rates) so we fix $n = 0.25$ and concentrate on the parameter $A$. In Figure \ref{fig:fig2}(d), we plot $A$ as a function of power for each excitation wavelength $\lambda$. We observe a clear trend towards rapid generation of large halos as the excitation energy increases, consistent with the photon energy exceeding the ground-state, single-photon ionization thresholds for both NV$^-$ and NV$^0$, that is, between 2.74 and 2.78\,eV ($\sim$450\,nm). 

The parameter $A$ informs the size of a halo at a given time and power. The data in Fig. \ref{fig:fig2}(d) is observed to follow a similar power law scaling as a function of optical power. Encapsulating the wavelength dependence into a single parameter $A'$ which we term the `activity', so that $r(\lambda,\tau, P) = A'(\lambda)t^{1/4}P^{n(\lambda)}$, we fit to the data in Fig. \ref{fig:fig2}(d) and extract $A'$ and $n$ as a function of wavelength. Fig. \ref{fig:fig2}(e) (inset) shows the extracted activity $A'$ ($n$) as a function of wavelength, for both ensemble-density samples considered in this work: sample A (nitrogen and NV only) and sample B (SiV, NV and nitrogen). For sample A we observe $n$ to vary from $n\approx 0.45$ at 633\,nm to $n \approx 0.25$ at 405\,nm, while for sample B $n$ remains nearly constant at $0.2-0.25$. Sample B, which has an overall higher defect concentration in addition to the presence of SiV centers admits smaller charge generation halos. This is consistent with an increased trap density (NV$^-$ and SiV$^{2-}$ for holes, N$^+$ for electrons) limiting the growth of the CCP, rather than any different charge generation processes (i.e. charge cycling of the SiV, which would be expected to result in larger CCP generation).   

From energetic considerations, we can determine three regimes spanned by the illumination energies used in this work. For $\lambda < 450\,$nm (2.76\,eV), ionization and recombination between NV charge states is mediated by single photon transitions from the defect's ground state. For $450 < \lambda < 575\,$nm (2.16 - 2.75\,eV), the excitation energy is sufficient to allow two-photon ionization and recombination processes, with the limit of $575\,$nm set by the ZPL of NV$^0$. This energy lower limit stems from the necessity to first excite the NV$^0$ $^2E\rightarrow^2A_2$ transition~\cite{manson_assignment_2013}, which promotes an $a_1$ electron to the $e_{x,y}$ state in the band gap, in so doing freeing a low-lying $a_1$ orbital in the band gap. Excitation of a valence electron to fill the band gap state is possible with photon energies as low as 1.2\,eV~\cite{meirzada_negative_2018}, completing the two-step process of photorecombination back to NV$^-$. For $\lambda > 575\,$nm, the charge cycling process is not clear due to the insufficient photon energy to first excite NV$^0$. We now study this observation more closely.

\section{Red charge cycling: phonon-assisted anti-Stokes driving}

With the observation that both orange (594\,nm) and red (633\,nm) light excites charge carrier generation in both SiV-rich and -poor samples, we conclude SiV centers play little if any role in cyclic carrier generation. Indeed, in the SiV-rich sample (B) the observed charge patterns are generally smaller and slower growing than in the NV-only sample (A). We therefore turn to the NV and nitrogen defects for an explanation. While carrier generation due to NV charge cycling is a reasonably well established process, the role of the substitutional nitrogen is not fully understood. For example, photoelectric measurements~\cite{bourgeois_enhanced_2017} and earlier bulk photoconductivity measurements~\cite{nesladek_dominant_1998, rosa_photoionization_1999} report a large background photocurrent assumed to originate from the nitrogen defects. While neutral nitrogen (N$^0$) occupies a donor level $1.7\,$eV (730\,nm) below the conduction band, the photoionization threshold has been measured to be closer to 2.2\,eV~\cite{nesladek_dominant_1998} due to the significant energy cost associated with rearranging the nitrogen and carbon atoms upon removal of the electron. While photoionization with red light is thus inefficient, this process generates only electrons, not holes, from N$^0\rightarrow$N$^+$, and direct photorecombination of N$^+\rightarrow$N$^0$ is forbidden due to the $\sim 4$\,eV acceptor level~\cite{jones_acceptor_2009}. 

\begin{figure}
	\centering
		\includegraphics[width = \columnwidth]{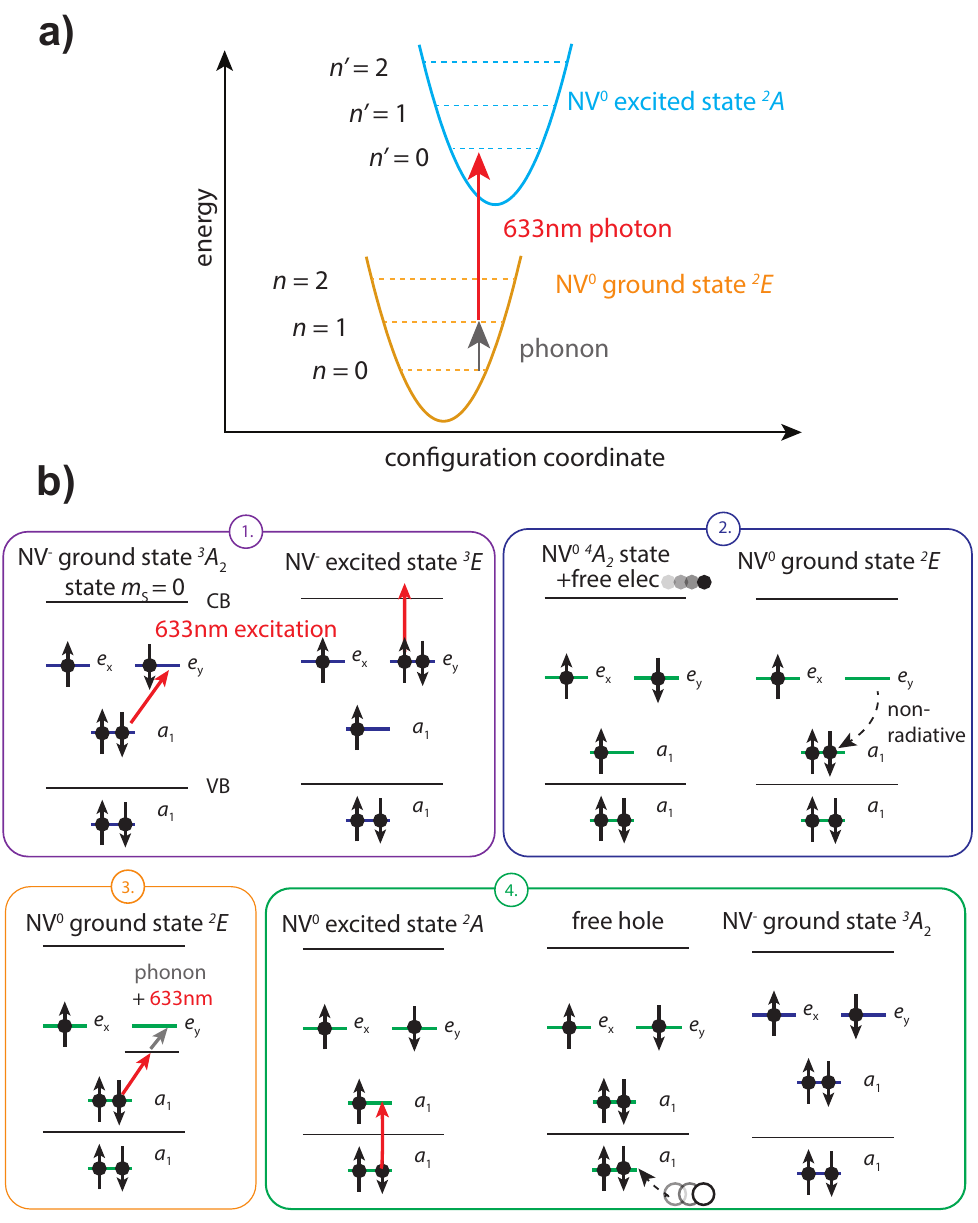}
	\caption{Anti-Stokes mediated NV charge cycling. a) Schematic diagram of phonon-assisted anti-Stokes excitation, where a photon with energy below the ZPL ($n'-n = 0$) and a phonon combine to drive an optical transition. Here, a phonon is first absorbed, which increases the vibrational quantum number $n$, followed by a phonon from the vibrationally excited state to the electronic excited state. b) Flow diagram in the single-electron picture of NV charge cycling under red light via phonon-assisted anti-Stokes excitation. (1) Starting in NV$^-$, sequential absorption of two red photons ionizes the NV. (2) Following Ref. \cite{razinkovas_photoionization_2021}, the NV is left in the metastable $^4A_2$ state of NV$^0$. We assume thence that non-radiative relaxation to the $^2E$ ground state occurs. (3) Phonon-assisted anti-Stokes absorption drives NV$^0$ into the $^2A$ excited state, thereby allowing recombination to NV$^-$ and generation of a free hole (4).}
	\label{fig:fig3}
\end{figure}

We therefore consider red-induced charge cycling of the NV directly, and posit phonon-assisted anti-Stokes excitation of the NV$^0$ $^2 E \rightarrow ^2 A$ transition as the cause. A schematic of the anti-Stokes process and charge cycling is depicted in Fig. \ref{fig:fig3}(a,b). In this process, a photon with insufficient energy (i.e. 633\,nm, 1.96\,eV) to directly excite an optical transition (the NV$^0$ ZPL, 2.16\,eV) is assisted by absorption of a phonon with sufficient energy to make up the difference in energy $\Delta E = 0.198\,$eV. Anti-Stokes excitation has been previously demonstrated for various color centers in wide band gap semiconductors~\cite{lin_anti-stokes_2022}. For diamond, specific pertinent examples being 720\,nm excitation of NV$^-$ ($\Delta E =0.224\,$eV)~\cite{tran_anti-stokes_2019}, 780\,nm excitation of SiV centers ($\Delta E = 0.09\,$eV)~\cite{gao_phonon-assisted_2018} (ZPL at 738\,nm) and recently for 594\,nm excitation of NV$^0$ centers ($\Delta E =0.07\,$eV)~\cite{gao_charge_2022}. Indeed, recent work examining wavelength-dependence of both photocurrent and photoelectrically-detected magnetic resonance (PDMR) measures both photocurrent and PDMR contrast persisting between 575\,nm and 637\,nm (where it terminates~\cite{todenhagen_wavelength_2023}), further evidence of NV charge cycling under red excitation. 

Two-photon ionization of NV$^-$ under red excitation is a well-understood process~\cite{aslam_photo-induced_2013}. However, recent work has questioned the electronic state of NV$^0$ immediately following ionization of NV$^-$~\cite{razinkovas_photoionization_2021} and the consequences for charge cycling have not been fully discussed to date. The process depicted in Fig. \ref{fig:fig3}(b) labeled (2) ascribes generation of NV$^0$ in the $^2 E$ ground state to non-radiative relaxation of the $^4A_2$ spin-quartet state that exists immediately following ionization, according to Ref. \cite{razinkovas_photoionization_2021}. Previous works had suggested an Auger process was required to repopulate the intra-band $a_1$ state following ionization from the $^3E$ state~\cite{siyushev_optically_2013, bourgeois_photoelectric_2020}. The post-ionization dynamics of the $^4A_2$ state were not considered in Ref. \cite{razinkovas_photoionization_2021}, and while described as metastable, the expected lifetime is of order 1\,$\upmu$s~\cite{gali_theory_2009}, making charge cycling slower but not exceptionally so. In principle, direct optical excitation of the $a_1$ valence band state to the $a_1$ state in the band gap could potentially occur during the time spent in the $^4A_2$ state, which skips the non-radiative relaxation step. While possible, this process cannot be the only means of charge cycling, as it requires an initial state every time of NV$^-$ for charge cycling to proceed. We see charge cycling proceed regardless of initialization into NV$^-$ or NV$^0$, with the latter assumed to be the equilibrium ground state of $^2E$.

\section{Modeling charge generation and capture}

For further understanding of the excitation wavelength dependence of charge carrier generation, we turn to a simulation of carrier generation and capture modeled by a system of equations that describe the time of evolution of hole, electron, NV centers and substitutional N concentrations in the presence of optical excitation~\cite{lozovoi_probing_2020, wood_room-temperature_2023}. Briefly, these equations track the time and space-dependent populations of NV and subsitutional nitrogen, with optically-generated electrons and holes assumed to follow a diffusive spatial expansion upon creation followed by capture by surrounding defects. More details are provided in Appendix A.

Given a hole generation rate $\Gamma$ and uniform NV$^-$ density $n$, one would simplistically assume that the radius of the CCP would increase with time as $r\sim (\Gamma t/\pi n)^{1/3}$, that is, the time required to fill a sphere of radius $r$. However, the scaling with optical power is not so easily extracted, and in any case the observed temporal dependence ($t^{1/4}$) deviates from this simple estimation. We were thus motivated to pursue a more in-depth analysis of carrier generation, diffusion and capture, not just of the NV centers but also the substitutional nitrogen, which has a strong effect on the diffusion of the carriers. 

We assume that the wavelength-dependent electron generation rate $\vartheta_-$ (ionization) and hole generation rate $\vartheta_+$ (recombination) are related to the optical power $P_0$ by
\begin{equation}
\vartheta_\pm(I,\lambda,r) = \left(\vartheta_{\pm, Q}(\lambda) P_0^2 + \vartheta_{\pm, L}(\lambda) P_0\right)e^{-2r^2/\sigma^2}
\label{eq:pp}
\end{equation} 
with the $Q,L$ subscripts denoting quadratic and linear terms and $\sigma$ the $1/e^2$ gaussian beam waist of the focused laser. For simplicity we focus only on Sample A, which contains only NV and N and ignore electron generation from ionization of N$^0$ and electron capture by NV$^0$, though we do include hole capture by N$^0$ as it is needed to fit the data. We assume radial symmetry, enabling the problem to be numerically solved one-dimensionally. We then vary the parameters $\vartheta_{\pm, L}$ and $\vartheta_{\pm, Q}$ to best fit the CCP radius-vs-power data presented in Fig. \ref{fig:fig2}(d) for Sample A. 

\begin{figure}
	\centering
		\includegraphics[width = \columnwidth]{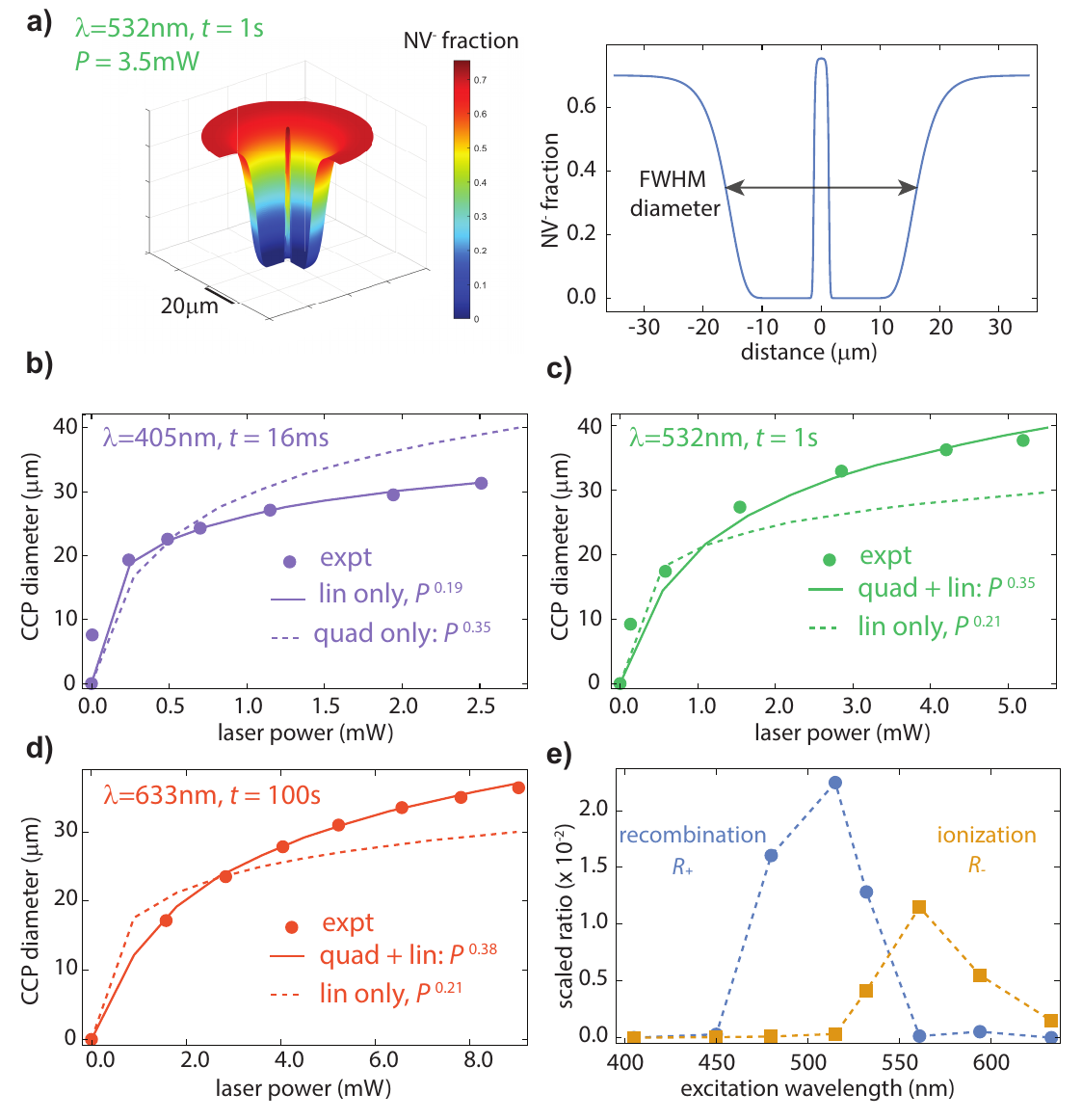}
	\caption{Numerical simulation of carrier diffusion and capture. a) left, Calculated NV$^-$ concentration as a function of position following a 1\,s park of a 3.5\,mW 532\,nm laser and right, lineout showing FWHM diameter of simulated CCP profile. The shape is in close agreement with experimental results (Fig. \ref{fig:fig2}(b)). b-d) Calculated CCP diameter as a function of laser power for $t = 16\,\text{ms}, \lambda = 405\,$nm (b), $t = 1\,\text{s}, \lambda = 532\,$nm (c) and $t = 100\,\text{s}, \lambda = 633\,$nm (d). Data points are experimental data, solid lines are best fits from optimization of ionization and recombination parameters, while dashed lines denote fits using only two-photon (b) or single-photon (c,d) power scaling terms. e) Scaled ratios of quadratic $\vartheta_Q$ and linear $\vartheta_L$ terms for ionization (blue) and recombination (orange) extracted from fits.}
	\label{fig:fig4}
\end{figure}

We obtain excellent correspondence between the theoretical and experimental results. For 405\,nm, both recombination and ionization are dominated by the one-photon processes, whereas in case of 532\,nm illumination, solely the two-photon processes are sufficient to reproduce the data, as expected from the single NV experiments. Our particular interest lies in understanding how the CCP grows as a function of power, given the power dependence offers insight into the underlying optical charge cycling mechanism. Our calculation results are presented in Fig. \ref{fig:fig4}. We define the characteristic diameter from a numerically-calculated CCP in an analogous manner to that used in experiments, as shown in Fig. \ref{fig:fig4}(a). We then show the fitted power dependence at fixed illumination times for three key excitation wavelengths, 405\,nm ($t = 16\,$ms), 532\,nm ($t = 1\,$s) and 633\,nm ($t = 100\,s$) in Fig. \ref{fig:fig4}(b-d). 

We are able to reproduce the characteristic temporal $\sim t^{0.25}$ and power $\sim P^{0.2n}$ dependence characteristic of the expansion of the CCP, and also find that as expected, explicit consideration of a mixture of one-photon ($\lambda <480\,$nm, $n = 1$) and two-photon processes ($n$=2) are key to accurately reproducing the experimental data. This latter point is best illustrated by the dashed lines in Fig. \ref{fig:fig4}(b-d), which are the fits including only two-photon power scaling (in the case of 405\,nm) and one-photon scaling (for 532\,nm and 633\,nm). Importantly, the growth of the charge capture halo for $\lambda> 575\,$nm is characteristic of a two-photon process, consistent with our hypothesis of anti-Stokes absorption mediating recombination of NV$^0$. Finally, we compare quadratic to linear ionization and recombination rates using the figure of merit 
\begin{equation}
R_\pm = \frac{\vartheta_{\pm, Q}}{\vartheta_{\pm, L} + 1},
\label{eq:ss}
\end{equation}  
defined so that when $\vartheta_{\pm, L}\rightarrow\infty$ then $R_\pm\rightarrow 0$, while for $\vartheta_{\pm, L}\rightarrow 0$ then $R_\pm\approx \vartheta_{\pm, Q}$, essentially quantifying when two-photon processes dominate over single-photon. We then calculate $R_\pm$ from extracted rates from numerical fits for each excitation wavelength and plot in Fig. \ref{fig:fig4}(e). Here, we see distinct profiles corresponding roughly to the absorption profiles of NV$^0$ (recombination) and NV$^-$ (ionization), evidence again of the underlying optical excitation processes driving CCP formation. These results offer an alternative to single-defect~\cite{aslam_photo-induced_2013} or photocurrent~\cite{todenhagen_wavelength_2023} spectroscopy to probe the optical dependence of charge carrier generation, and ultimately the level structure of unknown defects that charge cycle under optical excitation. 

\section{Charge transfer between single NVs}
\begin{figure*}
	\centering
		\includegraphics[width = \textwidth]{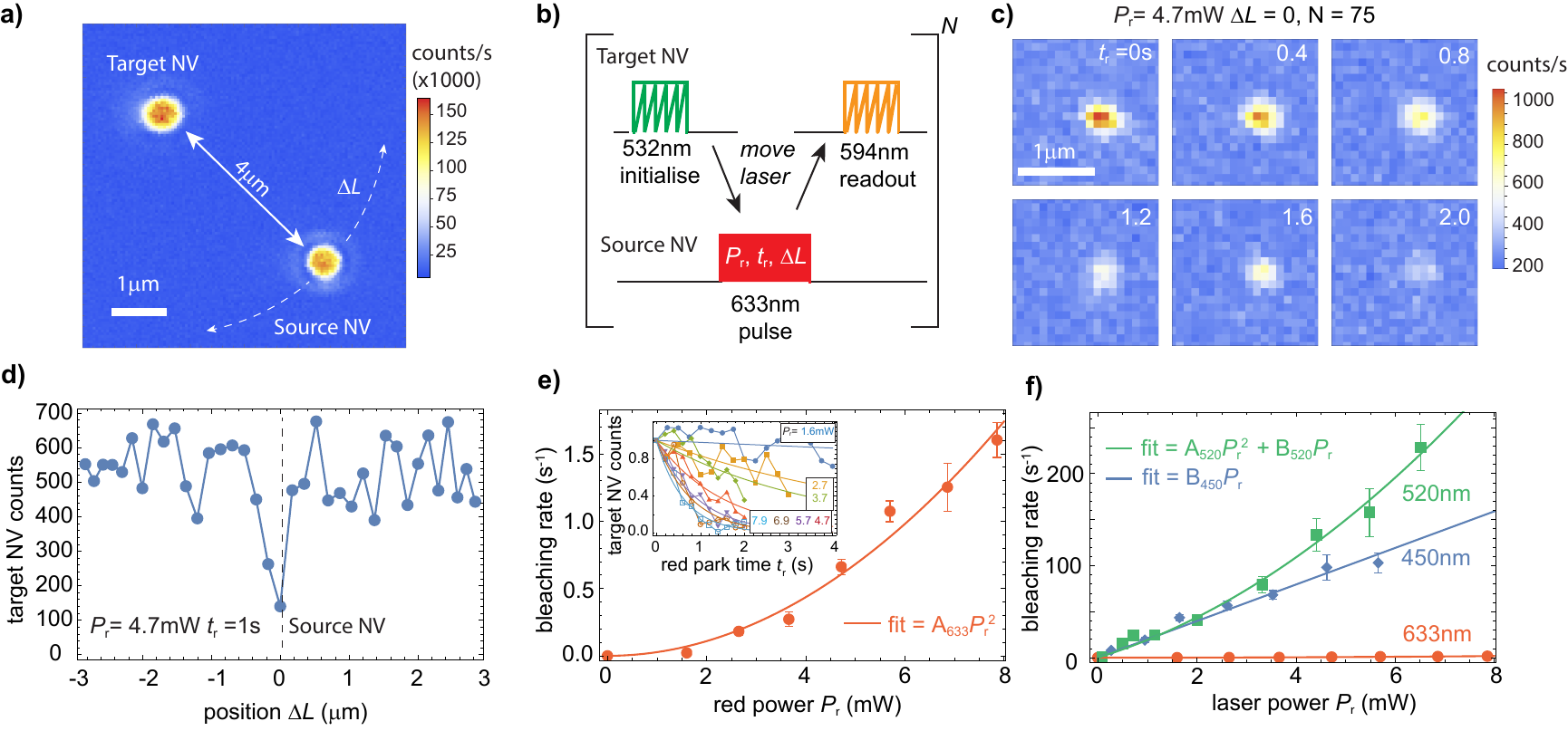}
	\caption{Red-induced charge transfer between two single NV centers. a) Two single NV centers in sample C, separated by 4$\,\upmu$m, denoted `source' and `target'. b) Optical protocol to initialize the target NV into the negative charge state with green light, followed by red illumination of the source NV to generate holes, which diffuse towards the target and are captured. A weak orange scan determines the charge state of the target, the whole procedure is repeated 75 times. c) Averaged 594\,nm confocal scans of the target NV following a 4.7\,mW red pulse on the source for varied times. The bleaching of NV$-$ fluorescence occurs because the target NV captures holes generated by the source NV, converting into the NV$^0$ state, and thus appears dark. d) Varying the position where the red laser is applied (shown in (a)) confirms red illumination of the source NV as the origin of the holes. e) Power dependence of bleaching rate reveals the characteristic quadratic dependence of a two-photon mediated recombination process at the source NV, (inset) time-variation of target NV charge state conversion as a function of the red laser power, solid lines denote fits of the form $e^{-\Gamma t_r}$, with $\Gamma$ parametrizing the bleaching rate.  f) Power dependence of red-induced hole generation compared to the much faster rates of green (520\,nm) and blue (450\,nm) light. Illumination with 450\,nm light results in single-photon mediated ionization/recombination, while 520\,nm light has both single- and two-photon components. Error bars are deduced from standard error in fitted parameters.}
	\label{fig:fig5}
\end{figure*}

To further corroborate our assertion that the NV center charge cycles slowly under red light, we turn to a third diamond (sample C) containing individually resolvable single NV centers. The sample has a low ($\sim$ppb) nitrogen concentration and sparse distribution of NV centers, approximately 10 single NV centers within a $(100\,\upmu\text{m})^2$, 10\,$\upmu$m-thick region of isotopically-enriched $^{12}$C (0.2\,\%), with a $\langle100\rangle$ surface. For these measurements, we switch to a second confocal microscope optimized for single-emitter detection, featuring an NA = 1.4 oil objective lens in an otherwise equivalent experimental configuration as described earlier. We identified two single NV centers (the `source', which will be illuminated by red light, and `target', which captures holes emitted from the source) separated by 4\,$\upmu$m, as shown in Fig. \ref{fig:fig5}(a), and implement the optical initialization and detection protocol depicted in Fig. \ref{fig:fig5}(b). 
Here, a green scan initializes the target NV into NV$^-$ with approximately 70$\%$ fidelity, no green initialization is performed on the source NV. The position of the laser focus is then moved to the source NV, which is illuminated with red (633\,nm) light with power $P_r$ for a duration $t_r$, with $\Delta L$ defining a $4\,\upmu$m radius arc where the red light position is varied, the source NV is located at $\Delta L = 0$. Following red pumping of the source, we move the laser focus back to the target and determine the charge state of the target with a 2\,$\upmu$W scan of 594\,nm light. With a photon energy below that of the NV$^0$ ZPL, 594\,nm orange light excites only NV$^-$ fluorescence for short illumination times, facilitating charge state detection~\cite{shields_efficient_2015}. A single orange light scan is insufficient to determine the charge state, so the whole sequence is repeated $N = 75$ times to generate averaged images, shown in Fig. \ref{fig:fig5}(c) for the case of a $P_r = 4.7\,$mW red laser pumping the target for varied times. The observed bleaching of NV$^-$ fluorescence is consistent with hole capture and formation of NV$^0$, which appears dark due to the selection of optical filters~\cite{lozovoi_optical_2021}.

While commonly regarded as a sub-micron, diffraction-limited spot in the $x$-$y$ imaging plane, the effective spatial extent of the illumination laser in our experiments is essentially a function of the power and illumination time. At high powers and illumination times approaching several seconds, the small fraction of power located in the Airy rings of the laser spot and various aberrations may result in direct bleaching of the target NV, even when the center of the beam is up to 2-3\,$\upmu$m away~\cite{gardill_super-resolution_2022}. We determined the effective $1/e^2$ width $w_e$ of the red laser beam by measuring the target NV bleaching as the laser was moved variable distances from the target along a line devoid of any NVs. For a 7\,mW, 10\,s exposure, $w_e =4.3(2)\,\upmu$m. Since the source NV is located $4\,\upmu$m away, provided the product of laser powers and illumination times are kept less than $70\,$mW\,s, the bleaching rate will not be affected by direct ionization of the target NV: the maximum we used in our measurements was 32\,mW\,s.

To further verify that the source NV is indeed the source of holes, we then varied the position of the red laser along a 4$\,\upmu$m radius arc with length $\Delta L$. In Figure \ref{fig:fig5}(d), for a 4.7\,mW, 1-s red pump on the source NV, we observed a narrow region centered on the position of the source NV ($\Delta L = 0$) where bleaching of the target NV is maximal. This confirms the source NV is the origin of the charge carriers captured by the target NV. We then examined the dependence of target NV bleaching on red power applied to the source NV, as shown in Fig. \ref{fig:fig5}(d). The target bleaching rate is observed to increase quadratically with red laser power, consistent with the hole generation at the source NV being mediated by a two-photon process. 

To put the slow charge cycling of NVs under red light into context, we then considered an equivalent demonstration of much more rapid charge cycling using blue (450\,nm) and green (520\,nm) light. With an identical procedure as outlined before, we measured the power dependence of hole generation and capture between source and target for blue and green light, as shown in Fig. \ref{fig:fig5}(e). For the same optical power, blue and green result in more than 100$\times$ faster hole generation. Indeed, for high blue or green excitation powers, hole capture by the target NV is detected within 1-5\,ms, on the order of the NV spin-lattice relaxation time $T_1\sim 5\,$ms. Extension to magnetic-resonance based charge carrier detection is discussed further in Section \ref{sec:disc}. 

The target-NV bleaching rates observed in this work are 10-20$\times$ faster than observed previously between single centers~\cite{lozovoi_optical_2021, lozovoi_detection_2023}. Since the bleaching rate is a combination of the generation rate (strongly dependent on wavelength and the specifics of a given optical setup), carrier transport rate (sample dependent) and NV$^-$ capture cross-section (assumed to be independent of sample), direct comparison is not necessarily straightforward. However, a key difference between this work and previous work is the method of NV center incorporation. While both samples are CVD-grown, electronic grade material (sample C slightly isotopically depleted at 0.2$\%$ $^{13}$C) and probed with similar experimental setups, the NV centers in our sample were incidentally incorporated during growth while those of Ref.~\cite{lozovoi_optical_2021, lozovoi_detection_2023} created with high-energy (20\,MeV) N$^+$ ion implantation. We observed similarly fast hole capture rates in another sample (data not shown) with a slightly higher NV density, though it too was an as-grown sample without any subsequent treatment.  

 While hole generation and bleaching with blue light has a clear linear dependence with power $(B_{450} = 19.8(8)\,\text{s}^{-1})$, a combined linear and quadratic model is required to fully explain the power dependence of the 520\,nm data $(A_{520} = 2.7(6)\,\text{s}^{-2}, B_{520} = 16.6\,\text{s}^{-1})$. Such a linear-quadratic dependence was discernible in previous work~\cite{lozovoi_optical_2021} though not significantly outside experimental uncertainties. Recently, an increase of photocurrent production under illumination with 520\,nm light was reported in NV ensembles~\cite{todenhagen_wavelength_2023} and attributed to a possible additional excited state of NV$^-$ in the conduction band, which may result in a more complicated, wavelength-dependent power dependence than a purely quadratic function. Regardless of additional excited states, the data in Fig. \ref{fig:fig2}(e) would imply hole generation with blue should be much faster than that of 520\,nm for the isolated single NVs. Here, we believe that imperfect aberration correction in the objective lens results in a slight position offset (notably in $z$) so that the laser spot size is not the same, nor in exactly the same position, for the two colors. Such an issue would not trouble our ensemble NV measurement, given the laser is always spatially overlapped with many NVs but can be severe for single NVs.  NV$^0$ fluorescence under 450\,nm excitation could be resolved, though was observed to have significantly degraded $z$-resolution. No NV fluorescence was observed under 405\,nm excitation.

\section{Discussion}\label{sec:disc}

In this work, we have examined how carrier photogeneration depends on illumination wavelength. Holes generated by NV charge cycling are captured by other NV$^-$, and the size of the region of NV$^0$ generated in a given time serves as a metric for the rate at which holes are generated. We studied diamonds with and without concentrations of silicon-vacancy centers, observing similar phenomenology that suggests SiV centers have a limited role in the generation of free holes in comparison to NV centers but affect the trasport of charge carriers through capture processes. At the very least, the contribution of NV centers to charge carrier generation under red illumination cannot be overlooked. We observe distinct regimes where differing mechanisms dominate the charge cycling rate, ranging from single-photon ionization for wavelengths $<450\,$nm, two-photon charge cycling for wavelengths between 450 and 561\,nm, and a weaker process via which photon energies below the zero-phonon line energy of NV$^0$ (2.15\,eV, 575\,nm) nevertheless manage to drive charge cycling (594 and 633\,nm). We suggest anti-Stokes excitation of the NV$^0$ $^2 E \rightarrow ^2 A_1$ transition as the process via which charge cycling is facilitated with red light. We confirm single NV centers charge cycle under red light this final regime by observing 633\,nm-induced charge transfer between two isolated single NV centers.

The principal outcome of this work is the confirmation that the charge state of the NV center cycles under red light. Though weak and correspondingly slow, red charge cycling is on the timescale of typical NV experimental averaging times. Our work has important consequences for the identification of charge generation processes in photoelectric detection measurements~\cite{hruby_magnetic_2022, todenhagen_wavelength_2023}, where the excitation energy is one of the few means available of identifying what defects contribute to photocurrent generation and efforts to understand the charge interconversion of other defects, such as SiV centers~\cite{dhomkar_-demand_2018, nicolas_optically_2019, gardill_probing_2021}. This latter point is of particular importance, since precision measurement of ionization and recombination thresholds is an essential means of confirming the accuracy of theoretical models~\cite{thiering_ab_2018}. Future work could confirm the hypothesis of phonon-assisted, anti-Stokes mediated charge cycling by studying the temperature dependence of either carrier generation and capture or photocurrent generation.

 An intriguing avenue of future inquiry concerns our final observation of rapid charge cycling and capture under blue and red light. For the highest powers used in these experiments, we could detect target NV hole capture and bleaching occurring in as little as 2\,ms. Such an interval is well within the T$_1$ time of the NVs we used and approaching the T$_2$ time of NVs in highly-isotopically-enriched $^{12}$C diamond ($\sim 1\,$ms). A future experiment could first initialize the target NV in the NV$^-$, $|m_S = 0\rangle$ state before inducing optical charge cycling on the proximal source NV. Microwave pulses could then drive the target NV into a superposition state and enact a magnetometry protocol -- such as a spin-echo sequence -- to detect how proximal charges affect the coherence of the NV spin. Direct detection of diffusing charges may even be possible by employing an electron-electron resonance scheme where a second microwave field drives the spins of the diffusing charge carriers. Hole generation and capture are readily observed optically via charge conversion, but the electrons generated by NV charge cycling remain hidden, making their role in capture processes with nonfluorescent defects (eg N$_S^+$) unclear~\cite{lozovoi_dark_2020, wang_manipulating_2023}. Spin-based detection schemes could begin to shed new light on these processes, and make advances towards proposed schemes of quantum information distribution at room temperature in diamond~\cite{doherty_towards_2016}. Simple additions such as higher laser powers or closer NVs could bring spin-coherent sensing of carrier generation closer, though hardware limitations impose a limit on how fast the laser can be moved from one place to another. A potential solution exists in the recently developed dual-beam confocal microscopy developed to enable correlation sensing~\cite{rovny_nanoscale_2022}.

In conclusion, we have experimentally examined the wavelength dependence of optically-driven charge cycling of the NV center in diamond, observing single-photon and two-photon driven charge cycling and hole capture in both ensembles of NVs as well as individual single centers. We have demonstrated that red light induces slow but measurable charge cycling of the NV center, which is unexpected given the discrepancy between the red photon energy (1.95\,eV) and NV$^0$ ZPL (2.16\,eV). We attribute this process to phonon-assisted anti-Stokes excitation of the NV$^0$ ZPL. With green and blue induced charge cycling of a single NV, we were able to demonstrate 100$\times$ faster charge cycling and single-center hole capture than under red light, and $10\times$ faster than reported previously, enabling carrier transport and capture within the $T_1$ time of each center. Our work will be important for the identification and characterization of defect charge states in diamond, photoelectric measurement of the NV center and could open the way to coherent spin-based measurements of carrier transport in a room-temperature, wide band-gap semiconductor.        

\section{Acknowledgments}
This work was supported by the Australian Research Council (DE210101093). R. M. G. was supported by an Australian Governement Research Training Program (RTP) Scholarship. A.L. acknowledges support from the National Science Foundation under grant NSF-2216838. C.A.M. acknowledges  support by the U.S. Department of Energy, Office of Science, National Quantum Information Science Research Centers, Co-design Center for Quantum Advantage (C2QA) under contract number DE-SC0012704. A.L. and C.A.M. acknowledge access to the facilities and research infrastructure of NSF CREST-IDEALS, grant number NSF-HRD-1547830.

\section*{Appendix A}
{\bf Carrier diffusion simulation details.}
We model the formation of halos upon photogeneration of charge carriers through a system of diffusion equations that describe the time and space evolution of NV$^-$ population $Q(\mathbf{r}, t)$, positively-charged substitutional nitrogen N$^+$ population $P(\mathbf{r}, t)$, electron density $n(\mathbf{r}, t)$ and hole density $p(\mathbf{r}, t)$. The equations are defined by admitting select processes that allow interconversion of NV and N charge states, which in turn generate electrons and holes that undergo diffusive transport, and may then be captured by more defects:
\begin{eqnarray}
\frac{\partial Q}{\partial t} & = & \left(Q_0 - Q\right) \vartheta_+ - Q \vartheta_-  - \sigma_{Qp}Q p\nonumber\\ 
\frac{\partial P}{\partial t} & = & - \sigma_{Ne}P n - \sigma_{Np}(P_0- P) p\nonumber\\ 
\frac{\partial n}{\partial t} & = & D_n \nabla^2 n + Q \vartheta_- - \sigma_{Ne}P n\nonumber\\ 
\frac{\partial p}{\partial t} & = & D_p \nabla^2 p -\sigma_{Qp}Q p - \sigma_{Np}(P_0- P) p, \\ 
\label{eq:}
\end{eqnarray} 
with $\vartheta_\pm$ as defined in Eq. \ref{eq:pp}, $Q_0$ the total volume concentration of NV (NV$^0$ and NV$^-$), $P_0$ the total volume concentration of N (N$^0$ and N$^+$), $D_n = 6.1\times10^9\,\upmu\text{m}^2/$s and $D_p = 5.3\times10^9\,\upmu\text{m}^2/$s the electron and hole diffusion coefficients in diamond~\cite{lozovoi_probing_2020}, and the coefficients $\sigma_{ab}$ representing probability of the capture of carrier $b$ by defect $a$, proportional to the cross-section for the given process and the carrier velocity: $\sigma_{Pn} \approx \sigma_{Qp} =  3.1\times10^5\,\upmu\text{m}^3/$s and $\sigma_{N^0 p}/\sigma_{Pn}\approx 10^{-2}$~\cite{lozovoi_probing_2020}. 

We assume that since the NV can charge cycle indefinitely under laser illumination, it is the source of all free carriers and ignore photoionization of the N$^0$ population. We used an NV concentration close to the value expected for the sample used for this work (10\,ppb), however the value used for N (3.4\,ppm) is higher than the 1\,ppm anticipated for Sample A. This was done to account for an unknown number of hole and electron traps – beyond NV and N – present in a crystal that significantly affect the halo growth.


\begin{thebibliography}{60}%
\makeatletter
\providecommand \@ifxundefined [1]{%
 \@ifx{#1\undefined}
}%
\providecommand \@ifnum [1]{%
 \ifnum #1\expandafter \@firstoftwo
 \else \expandafter \@secondoftwo
 \fi
}%
\providecommand \@ifx [1]{%
 \ifx #1\expandafter \@firstoftwo
 \else \expandafter \@secondoftwo
 \fi
}%
\providecommand \natexlab [1]{#1}%
\providecommand \enquote  [1]{``#1''}%
\providecommand \bibnamefont  [1]{#1}%
\providecommand \bibfnamefont [1]{#1}%
\providecommand \citenamefont [1]{#1}%
\providecommand \href@noop [0]{\@secondoftwo}%
\providecommand \href [0]{\begingroup \@sanitize@url \@href}%
\providecommand \@href[1]{\@@startlink{#1}\@@href}%
\providecommand \@@href[1]{\endgroup#1\@@endlink}%
\providecommand \@sanitize@url [0]{\catcode `\\12\catcode `\$12\catcode
  `\&12\catcode `\#12\catcode `\^12\catcode `\_12\catcode `\%12\relax}%
\providecommand \@@startlink[1]{}%
\providecommand \@@endlink[0]{}%
\providecommand \url  [0]{\begingroup\@sanitize@url \@url }%
\providecommand \@url [1]{\endgroup\@href {#1}{\urlprefix }}%
\providecommand \urlprefix  [0]{URL }%
\providecommand \Eprint [0]{\href }%
\providecommand \doibase [0]{https://doi.org/}%
\providecommand \selectlanguage [0]{\@gobble}%
\providecommand \bibinfo  [0]{\@secondoftwo}%
\providecommand \bibfield  [0]{\@secondoftwo}%
\providecommand \translation [1]{[#1]}%
\providecommand \BibitemOpen [0]{}%
\providecommand \bibitemStop [0]{}%
\providecommand \bibitemNoStop [0]{.\EOS\space}%
\providecommand \EOS [0]{\spacefactor3000\relax}%
\providecommand \BibitemShut  [1]{\csname bibitem#1\endcsname}%
\let\auto@bib@innerbib\@empty
%</preamble>
\bibitem [{\citenamefont {Doherty}\ \emph {et~al.}(2013)\citenamefont
  {Doherty}, \citenamefont {Manson}, \citenamefont {Delaney}, \citenamefont
  {Jelezko}, \citenamefont {Wrachtrup},\ and\ \citenamefont
  {Hollenberg}}]{doherty_nitrogen-vacancy_2013}%
  \BibitemOpen
  \bibfield  {author} {\bibinfo {author} {\bibfnamefont {M.~W.}\ \bibnamefont
  {Doherty}}, \bibinfo {author} {\bibfnamefont {N.~B.}\ \bibnamefont {Manson}},
  \bibinfo {author} {\bibfnamefont {P.}~\bibnamefont {Delaney}}, \bibinfo
  {author} {\bibfnamefont {F.}~\bibnamefont {Jelezko}}, \bibinfo {author}
  {\bibfnamefont {J.}~\bibnamefont {Wrachtrup}},\ and\ \bibinfo {author}
  {\bibfnamefont {L.~C.~L.}\ \bibnamefont {Hollenberg}},\ }\bibfield  {title}
  {\bibinfo {title} {The nitrogen-vacancy colour centre in diamond},\ }\href
  {https://doi.org/10.1016/j.physrep.2013.02.001} {\bibfield  {journal}
  {\bibinfo  {journal} {Physics Reports}},\ \textbf {\bibinfo {volume}
  {528}},\ \bibinfo {pages} {1} (\bibinfo {year} {2013})} \BibitemShut {NoStop}%
\bibitem [{\citenamefont {Thiering}\ and\ \citenamefont
  {Gali}(2020)}]{thiering_chapter_2020}%
  \BibitemOpen
  \bibfield  {author} {\bibinfo {author} {\bibfnamefont {G.}~\bibnamefont
  {Thiering}}\ and\ \bibinfo {author} {\bibfnamefont {A.}~\bibnamefont
  {Gali}},\ }\bibfield  {title} {\bibinfo {title} {{Color}
  centers in diamond for quantum applications},\ }in\ \href
  {https://doi.org/10.1016/bs.semsem.2020.03.001} {\emph {\bibinfo {booktitle}
  {Semiconductors and {Semimetals}}},\ \bibinfo {series} {Diamond for
  {Quantum} {Applications} {Part} 1}, Vol.\ \bibinfo {volume} {103}, pp.\ \bibinfo {pages}
  {1--36}},\ \bibinfo
  {editor} {eds.\ \bibinfo {editor} {\bibfnamefont {C.~E.}\ \bibnamefont
  {Nebel}}, \bibinfo {editor} {\bibfnamefont {I.}~\bibnamefont {Aharonovich}},
  \bibinfo {editor} {\bibfnamefont {N.}~\bibnamefont {Mizuochi}},\ and\
  \bibinfo {editor} {\bibfnamefont {M.}~\bibnamefont {Hatano}}}\ (\bibinfo
  {publisher} {Elsevier},\ \bibinfo {year} {2020})\BibitemShut {NoStop}%
\bibitem [{\citenamefont {Castelletto}\ and\ \citenamefont
  {Boretti}(2020)}]{castelletto_silicon_2020}%
  \BibitemOpen
  \bibfield  {author} {\bibinfo {author} {\bibfnamefont {S.}~\bibnamefont
  {Castelletto}}\ and\ \bibinfo {author} {\bibfnamefont {A.}~\bibnamefont
  {Boretti}},\ }\bibfield  {title} {\bibinfo {title} {Silicon carbide color
  centers for quantum applications},\ }\href
  {https://doi.org/10.1088/2515-7647/ab77a2} {\bibfield  {journal} {\bibinfo
  {journal} {J. Phys. Photonics}\ }\textbf {\bibinfo {volume} {2}},\ \bibinfo
  {pages} {022001} (\bibinfo {year} {2020})}\BibitemShut {NoStop}%
\bibitem [{\citenamefont {Tran}\ \emph {et~al.}(2016)\citenamefont {Tran},
  \citenamefont {Elbadawi}, \citenamefont {Totonjian}, \citenamefont {Lobo},
  \citenamefont {Grosso}, \citenamefont {Moon}, \citenamefont {Englund},
  \citenamefont {Ford}, \citenamefont {Aharonovich},\ and\ \citenamefont
  {Toth}}]{tran_robust_2016}%
  \BibitemOpen
  \bibfield  {author} {\bibinfo {author} {\bibfnamefont {T.~T.}\ \bibnamefont
  {Tran}}, \bibinfo {author} {\bibfnamefont {C.}~\bibnamefont {Elbadawi}},
  \bibinfo {author} {\bibfnamefont {D.}~\bibnamefont {Totonjian}}, \bibinfo
  {author} {\bibfnamefont {C.~J.}\ \bibnamefont {Lobo}}, \bibinfo {author}
  {\bibfnamefont {G.}~\bibnamefont {Grosso}}, \bibinfo {author} {\bibfnamefont
  {H.}~\bibnamefont {Moon}}, \bibinfo {author} {\bibfnamefont {D.~R.}\
  \bibnamefont {Englund}}, \bibinfo {author} {\bibfnamefont {M.~J.}\
  \bibnamefont {Ford}}, \bibinfo {author} {\bibfnamefont {I.}~\bibnamefont
  {Aharonovich}},\ and\ \bibinfo {author} {\bibfnamefont {M.}~\bibnamefont
  {Toth}},\ }\bibfield  {title} {\bibinfo {title} {Robust {Multicolor} {Single}
  {Photon} {Emission} from {Point} {Defects} in {Hexagonal} {Boron}
  {Nitride}},\ }\href {https://doi.org/10.1021/acsnano.6b03602} {\bibfield
  {journal} {\bibinfo  {journal} {ACS Nano}\ }\textbf {\bibinfo {volume}
  {10}},\ \bibinfo {pages} {7331} (\bibinfo {year} {2016})}\BibitemShut {NoStop}%
\bibitem [{\citenamefont {Caldwell}\ \emph {et~al.}(2019)\citenamefont
  {Caldwell}, \citenamefont {Aharonovich}, \citenamefont {Cassabois},
  \citenamefont {Edgar}, \citenamefont {Gil},\ and\ \citenamefont
  {Basov}}]{caldwell_photonics_2019}%
  \BibitemOpen
  \bibfield  {author} {\bibinfo {author} {\bibfnamefont {J.~D.}\ \bibnamefont
  {Caldwell}}, \bibinfo {author} {\bibfnamefont {I.}~\bibnamefont
  {Aharonovich}}, \bibinfo {author} {\bibfnamefont {G.}~\bibnamefont
  {Cassabois}}, \bibinfo {author} {\bibfnamefont {J.~H.}\ \bibnamefont
  {Edgar}}, \bibinfo {author} {\bibfnamefont {B.}~\bibnamefont {Gil}},\ and\
  \bibinfo {author} {\bibfnamefont {D.~N.}\ \bibnamefont {Basov}},\ }\bibfield
  {title} {\bibinfo {title} {Photonics with hexagonal boron nitride},\ }\href
  {https://doi.org/10.1038/s41578-019-0124-1} {\bibfield  {journal} {\bibinfo
  {journal} {Nat. Rev. Mater.}\ }\textbf {\bibinfo {volume} {4}},\ \bibinfo
  {pages} {552} (\bibinfo {year} {2019})}\BibitemShut {NoStop}%
\bibitem [{\citenamefont {Aslam}\ \emph {et~al.}(2013)\citenamefont {Aslam},
  \citenamefont {Waldherr}, \citenamefont {Neumann}, \citenamefont {Jelezko},\
  and\ \citenamefont {Wrachtrup}}]{aslam_photo-induced_2013}%
  \BibitemOpen
  \bibfield  {author} {\bibinfo {author} {\bibfnamefont {N.}~\bibnamefont
  {Aslam}}, \bibinfo {author} {\bibfnamefont {G.}~\bibnamefont {Waldherr}},
  \bibinfo {author} {\bibfnamefont {P.}~\bibnamefont {Neumann}}, \bibinfo
  {author} {\bibfnamefont {F.}~\bibnamefont {Jelezko}},\ and\ \bibinfo {author}
  {\bibfnamefont {J.}~\bibnamefont {Wrachtrup}},\ }\bibfield  {title} {\bibinfo
  {title} {Photo-induced ionization dynamics of the nitrogen vacancy defect in
  diamond investigated by single-shot charge state detection},\ }\href
  {https://doi.org/10.1088/1367-2630/15/1/013064} {\bibfield  {journal}
  {\bibinfo  {journal} {New J. Phys.}\ }\textbf {\bibinfo {volume} {15}},\
  \bibinfo {pages} {013064} (\bibinfo {year} {2013})}\BibitemShut {NoStop}%
\bibitem [{\citenamefont {Waldherr}\ \emph {et~al.}(2011)\citenamefont
  {Waldherr}, \citenamefont {Beck}, \citenamefont {Steiner}, \citenamefont
  {Neumann}, \citenamefont {Gali}, \citenamefont {Frauenheim}, \citenamefont
  {Jelezko},\ and\ \citenamefont {Wrachtrup}}]{waldherr_dark_2011}%
  \BibitemOpen
  \bibfield  {author} {\bibinfo {author} {\bibfnamefont {G.}~\bibnamefont
  {Waldherr}}, \bibinfo {author} {\bibfnamefont {J.}~\bibnamefont {Beck}},
  \bibinfo {author} {\bibfnamefont {M.}~\bibnamefont {Steiner}}, \bibinfo
  {author} {\bibfnamefont {P.}~\bibnamefont {Neumann}}, \bibinfo {author}
  {\bibfnamefont {A.}~\bibnamefont {Gali}}, \bibinfo {author} {\bibfnamefont
  {T.}~\bibnamefont {Frauenheim}}, \bibinfo {author} {\bibfnamefont
  {F.}~\bibnamefont {Jelezko}},\ and\ \bibinfo {author} {\bibfnamefont
  {J.}~\bibnamefont {Wrachtrup}},\ }\bibfield  {title} {\bibinfo {title} {Dark
  {States} of {Single} {Nitrogen}-{Vacancy} {Centers} in {Diamond} {Unraveled}
  by {Single} {Shot} {NMR}},\ }\href
  {https://doi.org/10.1103/PhysRevLett.106.157601} {\bibfield  {journal}
  {\bibinfo  {journal} {Phys. Rev. Lett.}\ }\textbf {\bibinfo {volume} {106}},\
  \bibinfo {pages} {157601} (\bibinfo {year} {2011})}\BibitemShut {NoStop}%
\bibitem [{\citenamefont {Rose}\ \emph {et~al.}(2018)\citenamefont {Rose},
  \citenamefont {Huang}, \citenamefont {Zhang}, \citenamefont {Stevenson},
  \citenamefont {Tyryshkin}, \citenamefont {Sangtawesin}, \citenamefont
  {Srinivasan}, \citenamefont {Loudin}, \citenamefont {Markham}, \citenamefont
  {Edmonds}, \citenamefont {Twitchen}, \citenamefont {Lyon},\ and\
  \citenamefont {de~Leon}}]{rose_observation_2018}%
  \BibitemOpen
  \bibfield  {author} {\bibinfo {author} {\bibfnamefont {B.~C.}\ \bibnamefont
  {Rose}}, \bibinfo {author} {\bibfnamefont {D.}~\bibnamefont {Huang}},
  \bibinfo {author} {\bibfnamefont {Z.-H.}\ \bibnamefont {Zhang}}, \bibinfo
  {author} {\bibfnamefont {P.}~\bibnamefont {Stevenson}}, \bibinfo {author}
  {\bibfnamefont {A.~M.}\ \bibnamefont {Tyryshkin}}, \bibinfo {author}
  {\bibfnamefont {S.}~\bibnamefont {Sangtawesin}}, \bibinfo {author}
  {\bibfnamefont {S.}~\bibnamefont {Srinivasan}}, \bibinfo {author}
  {\bibfnamefont {L.}~\bibnamefont {Loudin}}, \bibinfo {author} {\bibfnamefont
  {M.~L.}\ \bibnamefont {Markham}}, \bibinfo {author} {\bibfnamefont {A.~M.}\
  \bibnamefont {Edmonds}}, \bibinfo {author} {\bibfnamefont {D.~J.}\
  \bibnamefont {Twitchen}}, \bibinfo {author} {\bibfnamefont {S.~A.}\
  \bibnamefont {Lyon}},\ and\ \bibinfo {author} {\bibfnamefont {N.~P.}\
  \bibnamefont {de~Leon}},\ }\bibfield  {title} {\bibinfo {title} {Observation
  of an environmentally insensitive solid-state spin defect in diamond},\
  }\href {https://doi.org/10.1126/science.aao0290} {\bibfield  {journal}
  {\bibinfo  {journal} {Science}\ }\textbf {\bibinfo {volume} {361}},\ \bibinfo
  {pages} {60} (\bibinfo {year} {2018})}\BibitemShut {NoStop}%
\bibitem [{\citenamefont {Wood}\ \emph {et~al.}(2023)\citenamefont {Wood},
  \citenamefont {Lozovoi}, \citenamefont {Zhang}, \citenamefont {Sharma},
  \citenamefont {López-Morales}, \citenamefont {Jayakumar}, \citenamefont
  {de~Leon},\ and\ \citenamefont {Meriles}}]{wood_room-temperature_2023}%
  \BibitemOpen
  \bibfield  {author} {\bibinfo {author} {\bibfnamefont {A.}~\bibnamefont
  {Wood}}, \bibinfo {author} {\bibfnamefont {A.}~\bibnamefont {Lozovoi}},
  \bibinfo {author} {\bibfnamefont {Z.-H.}\ \bibnamefont {Zhang}}, \bibinfo
  {author} {\bibfnamefont {S.}~\bibnamefont {Sharma}}, \bibinfo {author}
  {\bibfnamefont {G.~I.}\ \bibnamefont {López-Morales}}, \bibinfo {author}
  {\bibfnamefont {H.}~\bibnamefont {Jayakumar}}, \bibinfo {author}
  {\bibfnamefont {N.~P.}\ \bibnamefont {de~Leon}},\ and\ \bibinfo {author}
  {\bibfnamefont {C.~A.}\ \bibnamefont {Meriles}},\ }\bibfield  {title}
  {\bibinfo {title} {Room-{Temperature} {Photochromism} of {Silicon} {Vacancy}
  {Centers} in {CVD} {Diamond}},\ }\href
  {https://doi.org/10.1021/acs.nanolett.2c04514} {\bibfield  {journal}
  {\bibinfo  {journal} {Nano Lett.}\ }\textbf {\bibinfo {volume} {23}},\
  \bibinfo {pages} {1017} (\bibinfo {year} {2023})}\BibitemShut {NoStop}%
\bibitem [{\citenamefont {Zhang}\ \emph {et~al.}(2023)\citenamefont {Zhang},
  \citenamefont {Edmonds}, \citenamefont {Palmer}, \citenamefont {Markham},\
  and\ \citenamefont {de~Leon}}]{zhang_neutral_2023}%
  \BibitemOpen
  \bibfield  {author} {\bibinfo {author} {\bibfnamefont {Z.-H.}\ \bibnamefont
  {Zhang}}, \bibinfo {author} {\bibfnamefont {A.~M.}\ \bibnamefont {Edmonds}},
  \bibinfo {author} {\bibfnamefont {N.}~\bibnamefont {Palmer}}, \bibinfo
  {author} {\bibfnamefont {M.~L.}\ \bibnamefont {Markham}},\ and\ \bibinfo
  {author} {\bibfnamefont {N.~P.}\ \bibnamefont {de~Leon}},\ }\bibfield
  {title} {\bibinfo {title} {Neutral {Silicon}-{Vacancy} {Centers} in {Diamond}
  via {Photoactivated} {Itinerant} {Carriers}},\ }\href
  {https://doi.org/10.1103/PhysRevApplied.19.034022} {\bibfield  {journal}
  {\bibinfo  {journal} {Phys. Rev. Appl.}\ }\textbf {\bibinfo {volume} {19}},\
  \bibinfo {pages} {034022} (\bibinfo {year} {2023})}\BibitemShut {NoStop}%
\bibitem [{\citenamefont {Ashfold}\ \emph {et~al.}(2020)\citenamefont
  {Ashfold}, \citenamefont {Goss}, \citenamefont {Green}, \citenamefont {May},
  \citenamefont {Newton},\ and\ \citenamefont
  {Peaker}}]{ashfold_nitrogen_2020}%
  \BibitemOpen
  \bibfield  {author} {\bibinfo {author} {\bibfnamefont {M.~N.~R.}\
  \bibnamefont {Ashfold}}, \bibinfo {author} {\bibfnamefont {J.~P.}\
  \bibnamefont {Goss}}, \bibinfo {author} {\bibfnamefont {B.~L.}\ \bibnamefont
  {Green}}, \bibinfo {author} {\bibfnamefont {P.~W.}\ \bibnamefont {May}},
  \bibinfo {author} {\bibfnamefont {M.~E.}\ \bibnamefont {Newton}},\ and\
  \bibinfo {author} {\bibfnamefont {C.~V.}\ \bibnamefont {Peaker}},\ }\bibfield
   {title} {\bibinfo {title} {Nitrogen in {Diamond}},\ }\href
  {https://doi.org/10.1021/acs.chemrev.9b00518} {\bibfield  {journal} {\bibinfo
   {journal} {Chem. Rev.}\ }\textbf {\bibinfo {volume} {120}},\ \bibinfo
  {pages} {5745} (\bibinfo {year} {2020})}\BibitemShut {NoStop}%
\bibitem [{\citenamefont {Manson}\ \emph {et~al.}(2018)\citenamefont {Manson},
  \citenamefont {Hedges}, \citenamefont {S}, \citenamefont {Barson},
  \citenamefont {Ahlefeldt}, \citenamefont {Doherty}, \citenamefont {Sellars},
  \citenamefont {Abe},\ and\ \citenamefont {Ohshima}}]{manson_NV-_2018}%
  \BibitemOpen
  \bibfield  {author} {\bibinfo {author} {\bibfnamefont {N.~B.}\ \bibnamefont
  {Manson}}, \bibinfo {author} {\bibfnamefont {M.}~\bibnamefont {Hedges}},
  \bibinfo {author} {\bibfnamefont {M.}~\bibnamefont {S}}, \bibinfo {author}
  {\bibfnamefont {J.}~\bibnamefont {Barson}}, \bibinfo {author} {\bibfnamefont
  {R.}~\bibnamefont {Ahlefeldt}}, \bibinfo {author} {\bibfnamefont {M.~W.}\
  \bibnamefont {Doherty}}, \bibinfo {author} {\bibfnamefont {M.~J.}\
  \bibnamefont {Sellars}}, \bibinfo {author} {\bibfnamefont {H.}~\bibnamefont
  {Abe}},\ and\ \bibinfo {author} {\bibfnamefont {T.}~\bibnamefont {Ohshima}},\
  }\bibfield  {title} {\bibinfo {title} {{NV}- - {N}+ pair centre in 1b
  diamond},\ }\href {https://doi.org/10.1088/1367-2630/aaec58} {\bibfield
  {journal} {\bibinfo  {journal} {New J. Phys.}\ }\textbf {\bibinfo {volume}
  {20}},\ \bibinfo {pages} {113037} (\bibinfo {year} {2018})}\BibitemShut {NoStop}%
\bibitem [{\citenamefont {Lozovoi}\ \emph
  {et~al.}(2020{\natexlab{a}})\citenamefont {Lozovoi}, \citenamefont {Daw},
  \citenamefont {Jayakumar},\ and\ \citenamefont
  {Meriles}}]{lozovoi_dark_2020}%
  \BibitemOpen
  \bibfield  {author} {\bibinfo {author} {\bibfnamefont {A.}~\bibnamefont
  {Lozovoi}}, \bibinfo {author} {\bibfnamefont {D.}~\bibnamefont {Daw}},
  \bibinfo {author} {\bibfnamefont {H.}~\bibnamefont {Jayakumar}},\ and\
  \bibinfo {author} {\bibfnamefont {C.~A.}\ \bibnamefont {Meriles}},\
  }\bibfield  {title} {\bibinfo {title} {Dark defect charge dynamics in bulk
  chemical-vapor-deposition-grown diamonds probed via nitrogen vacancy
  centers},\ }\href {https://doi.org/10.1103/PhysRevMaterials.4.053602}
  {\bibfield  {journal} {\bibinfo  {journal} {Phys. Rev. Materials}\ }\textbf
  {\bibinfo {volume} {4}},\ \bibinfo {pages} {053602} (\bibinfo {year}
  {2020}{\natexlab{a}})}\BibitemShut {NoStop}%
\bibitem [{\citenamefont {Wang}\ \emph {et~al.}(2023)\citenamefont {Wang},
  \citenamefont {Li}, \citenamefont {Tang}, \citenamefont {Li}, \citenamefont
  {Madonini}, \citenamefont {Alsallom}, \citenamefont {Sun}, \citenamefont
  {Peng}, \citenamefont {Villa}, \citenamefont {Li},\ and\ \citenamefont
  {Cappellaro}}]{wang_manipulating_2023}%
  \BibitemOpen
  \bibfield  {author} {\bibinfo {author} {\bibfnamefont {G.}~\bibnamefont
  {Wang}}, \bibinfo {author} {\bibfnamefont {C.}~\bibnamefont {Li}}, \bibinfo
  {author} {\bibfnamefont {H.}~\bibnamefont {Tang}}, \bibinfo {author}
  {\bibfnamefont {B.}~\bibnamefont {Li}}, \bibinfo {author} {\bibfnamefont
  {F.}~\bibnamefont {Madonini}}, \bibinfo {author} {\bibfnamefont {F.~F.}\
  \bibnamefont {Alsallom}}, \bibinfo {author} {\bibfnamefont {W.~K.~C.}\
  \bibnamefont {Sun}}, \bibinfo {author} {\bibfnamefont {P.}~\bibnamefont
  {Peng}}, \bibinfo {author} {\bibfnamefont {F.}~\bibnamefont {Villa}},
  \bibinfo {author} {\bibfnamefont {J.}~\bibnamefont {Li}},\ and\ \bibinfo
  {author} {\bibfnamefont {P.}~\bibnamefont {Cappellaro}},} \bibinfo {title} {Manipulating
  solid-state spin concentration through charge transport}\href
  {https://doi.org/10.48550/arXiv.2302.12742} {\bibinfo {note} {arXiv:2302.12742}(\bibinfo {year}
  {2023})} \BibitemShut {NoStop}%
\bibitem [{\citenamefont {Barry}\ \emph {et~al.}(2020)\citenamefont {Barry},
  \citenamefont {Schloss}, \citenamefont {Bauch}, \citenamefont {Turner},
  \citenamefont {Hart}, \citenamefont {Pham},\ and\ \citenamefont
  {Walsworth}}]{barry_sensitivity_2020}%
  \BibitemOpen
  \bibfield  {author} {\bibinfo {author} {\bibfnamefont {J.~F.}\ \bibnamefont
  {Barry}}, \bibinfo {author} {\bibfnamefont {J.~M.}\ \bibnamefont {Schloss}},
  \bibinfo {author} {\bibfnamefont {E.}~\bibnamefont {Bauch}}, \bibinfo
  {author} {\bibfnamefont {M.~J.}\ \bibnamefont {Turner}}, \bibinfo {author}
  {\bibfnamefont {C.~A.}\ \bibnamefont {Hart}}, \bibinfo {author}
  {\bibfnamefont {L.~M.}\ \bibnamefont {Pham}},\ and\ \bibinfo {author}
  {\bibfnamefont {R.~L.}\ \bibnamefont {Walsworth}},\ }\bibfield  {title}
  {\bibinfo {title} {Sensitivity optimization for {NV}-diamond magnetometry},\
  }\href {https://doi.org/10.1103/RevModPhys.92.015004} {\bibfield  {journal}
  {\bibinfo  {journal} {Rev. Mod. Phys.}\ }\textbf {\bibinfo {volume} {92}},\
  \bibinfo {pages} {015004} (\bibinfo {year} {2020})}\BibitemShut {NoStop}%
\bibitem [{\citenamefont {Han}\ \emph {et~al.}(2010)\citenamefont {Han},
  \citenamefont {Kim}, \citenamefont {Eggeling},\ and\ \citenamefont
  {Hell}}]{han_metastable_2010}%
  \BibitemOpen
  \bibfield  {author} {\bibinfo {author} {\bibfnamefont {K.~Y.}\ \bibnamefont
  {Han}}, \bibinfo {author} {\bibfnamefont {S.~K.}\ \bibnamefont {Kim}},
  \bibinfo {author} {\bibfnamefont {C.}~\bibnamefont {Eggeling}},\ and\
  \bibinfo {author} {\bibfnamefont {S.~W.}\ \bibnamefont {Hell}},\ }\bibfield
  {title} {\bibinfo {title} {Metastable {Dark} {States} {Enable} {Ground}
  {State} {Depletion} {Microscopy} of {Nitrogen} {Vacancy} {Centers} in
  {Diamond} with {Diffraction}-{Unlimited} {Resolution}},\ }\href
  {https://doi.org/10.1021/nl102156m} {\bibfield  {journal} {\bibinfo
  {journal} {Nano Lett.}\ }\textbf {\bibinfo {volume} {10}},\ \bibinfo {pages}
  {3199} (\bibinfo {year} {2010})}\BibitemShut {NoStop}%
\bibitem [{\citenamefont {Shields}\ \emph {et~al.}(2015)\citenamefont
  {Shields}, \citenamefont {Unterreithmeier}, \citenamefont {de~Leon},
  \citenamefont {Park},\ and\ \citenamefont {Lukin}}]{shields_efficient_2015}%
  \BibitemOpen
  \bibfield  {author} {\bibinfo {author} {\bibfnamefont {B.}~\bibnamefont
  {Shields}}, \bibinfo {author} {\bibfnamefont {Q.}~\bibnamefont
  {Unterreithmeier}}, \bibinfo {author} {\bibfnamefont {N.}~\bibnamefont
  {de~Leon}}, \bibinfo {author} {\bibfnamefont {H.}~\bibnamefont {Park}},\ and\
  \bibinfo {author} {\bibfnamefont {M.}~\bibnamefont {Lukin}},\ }\bibfield
  {title} {\bibinfo {title} {Efficient {Readout} of a {Single} {Spin} {State}
  in {Diamond} via {Spin}-to-{Charge} {Conversion}},\ }\href
  {https://doi.org/10.1103/PhysRevLett.114.136402} {\bibfield  {journal}
  {\bibinfo  {journal} {Phys. Rev. Lett.}\ }\textbf {\bibinfo {volume} {114}},\
  \bibinfo {pages} {136402} (\bibinfo {year} {2015})}\BibitemShut {NoStop}%
\bibitem [{\citenamefont {Dhomkar}\ \emph {et~al.}(2016)\citenamefont
  {Dhomkar}, \citenamefont {Henshaw}, \citenamefont {Jayakumar},\ and\
  \citenamefont {Meriles}}]{dhomkar_long-term_2016}%
  \BibitemOpen
  \bibfield  {author} {\bibinfo {author} {\bibfnamefont {S.}~\bibnamefont
  {Dhomkar}}, \bibinfo {author} {\bibfnamefont {J.}~\bibnamefont {Henshaw}},
  \bibinfo {author} {\bibfnamefont {H.}~\bibnamefont {Jayakumar}},\ and\
  \bibinfo {author} {\bibfnamefont {C.~A.}\ \bibnamefont {Meriles}},\
  }\bibfield  {title} {\bibinfo {title} {Long-term data storage in diamond},\
  }\href {https://doi.org/10.1126/sciadv.1600911} {\bibfield  {journal}
  {\bibinfo  {journal} {Sci. Adv.}\ }\textbf {\bibinfo {volume} {2}},\
  \bibinfo {pages} {e1600911} (\bibinfo {year} {2016})}\BibitemShut
  {NoStop}%
\bibitem [{\citenamefont {Pfender}\ \emph {et~al.}(2017)\citenamefont
  {Pfender}, \citenamefont {Aslam}, \citenamefont {Simon}, \citenamefont
  {Antonov}, \citenamefont {Thiering}, \citenamefont {Burk}, \citenamefont
  {Fávaro~de Oliveira}, \citenamefont {Denisenko}, \citenamefont {Fedder},
  \citenamefont {Meijer}, \citenamefont {Garrido}, \citenamefont {Gali},
  \citenamefont {Teraji}, \citenamefont {Isoya}, \citenamefont {Doherty},
  \citenamefont {Alkauskas}, \citenamefont {Gallo}, \citenamefont {Grüneis},
  \citenamefont {Neumann},\ and\ \citenamefont
  {Wrachtrup}}]{pfender_protecting_2017}%
  \BibitemOpen
  \bibfield  {author} {\bibinfo {author} {\bibfnamefont {M.}~\bibnamefont
  {Pfender}}, \bibinfo {author} {\bibfnamefont {N.}~\bibnamefont {Aslam}},
  \bibinfo {author} {\bibfnamefont {P.}~\bibnamefont {Simon}}, \bibinfo
  {author} {\bibfnamefont {D.}~\bibnamefont {Antonov}}, \bibinfo {author}
  {\bibfnamefont {G.}~\bibnamefont {Thiering}}, \bibinfo {author}
  {\bibfnamefont {S.}~\bibnamefont {Burk}}, \bibinfo {author} {\bibfnamefont
  {F.}~\bibnamefont {Fávaro~de Oliveira}}, \bibinfo {author} {\bibfnamefont
  {A.}~\bibnamefont {Denisenko}}, \bibinfo {author} {\bibfnamefont
  {H.}~\bibnamefont {Fedder}}, \bibinfo {author} {\bibfnamefont
  {J.}~\bibnamefont {Meijer}}, \bibinfo {author} {\bibfnamefont {J.~A.}\
  \bibnamefont {Garrido}}, \bibinfo {author} {\bibfnamefont {A.}~\bibnamefont
  {Gali}}, \bibinfo {author} {\bibfnamefont {T.}~\bibnamefont {Teraji}},
  \bibinfo {author} {\bibfnamefont {J.}~\bibnamefont {Isoya}}, \bibinfo
  {author} {\bibfnamefont {M.~W.}\ \bibnamefont {Doherty}}, \bibinfo {author}
  {\bibfnamefont {A.}~\bibnamefont {Alkauskas}}, \bibinfo {author}
  {\bibfnamefont {A.}~\bibnamefont {Gallo}}, \bibinfo {author} {\bibfnamefont
  {A.}~\bibnamefont {Grüneis}}, \bibinfo {author} {\bibfnamefont
  {P.}~\bibnamefont {Neumann}},\ and\ \bibinfo {author} {\bibfnamefont
  {J.}~\bibnamefont {Wrachtrup}},\ }\bibfield  {title} {\bibinfo {title}
  {Protecting a {Diamond} {Quantum} {Memory} by {Charge} {State} {Control}},\
  }\href {https://doi.org/10.1021/acs.nanolett.7b01796} {\bibfield  {journal}
  {\bibinfo  {journal} {Nano Lett.}\ }\textbf {\bibinfo {volume} {17}},\
  \bibinfo {pages} {5931} (\bibinfo {year} {2017})}\BibitemShut {NoStop}%
\bibitem [{\citenamefont {McCloskey}\ \emph {et~al.}(2022)\citenamefont
  {McCloskey}, \citenamefont {Dontschuk}, \citenamefont {Stacey}, \citenamefont
  {Pattinson}, \citenamefont {Nadarajah}, \citenamefont {Hall}, \citenamefont
  {Hollenberg}, \citenamefont {Prawer},\ and\ \citenamefont
  {Simpson}}]{mccloskey_diamond_2022}%
  \BibitemOpen
  \bibfield  {author} {\bibinfo {author} {\bibfnamefont {D.~J.}\ \bibnamefont
  {McCloskey}}, \bibinfo {author} {\bibfnamefont {N.}~\bibnamefont
  {Dontschuk}}, \bibinfo {author} {\bibfnamefont {A.}~\bibnamefont {Stacey}},
  \bibinfo {author} {\bibfnamefont {C.}~\bibnamefont {Pattinson}}, \bibinfo
  {author} {\bibfnamefont {A.}~\bibnamefont {Nadarajah}}, \bibinfo {author}
  {\bibfnamefont {L.~T.}\ \bibnamefont {Hall}}, \bibinfo {author}
  {\bibfnamefont {L.~C.~L.}\ \bibnamefont {Hollenberg}}, \bibinfo {author}
  {\bibfnamefont {S.}~\bibnamefont {Prawer}},\ and\ \bibinfo {author}
  {\bibfnamefont {D.~A.}\ \bibnamefont {Simpson}},\ }\bibfield  {title}
  {\bibinfo {title} {A diamond voltage imaging microscope},\ }\href
  {https://doi.org/10.1038/s41566-022-01064-1} {\bibfield  {journal} {\bibinfo
  {journal} {Nat. Photon.}\ }\textbf {\bibinfo {volume} {16}},\ \bibinfo
  {pages} {730} (\bibinfo {year} {2022})}\BibitemShut {NoStop}%
\bibitem [{\citenamefont {Monge}\ \emph {et~al.}(2023)\citenamefont {Monge},
  \citenamefont {Delord},\ and\ \citenamefont
  {Meriles}}]{monge_reversible_2023}%
  \BibitemOpen
  \bibfield  {author} {\bibinfo {author} {\bibfnamefont {R.}~\bibnamefont
  {Monge}}, \bibinfo {author} {\bibfnamefont {T.}~\bibnamefont {Delord}},\ and\
  \bibinfo {author} {\bibfnamefont {C.~A.}\ \bibnamefont {Meriles}},\
  }\bibfield  {title} {\bibinfo {title} {Reversible optical data storage below
  the diffraction limit},\ }\href {https://doi.org/10.1038/s41565-023-01542-9}
  {\bibfield  {journal} {\bibinfo  {journal} {Nat. Nanotechnol.}\ ,\ \bibinfo
  {pages} {1}} (\bibinfo {year} {2023})}\BibitemShut {NoStop}%
\bibitem [{\citenamefont {Ji}\ \emph {et~al.}(2024)\citenamefont {Ji},
  \citenamefont {Liu}, \citenamefont {Guo}, \citenamefont {Hu}, \citenamefont
  {Zhou}, \citenamefont {Dai}, \citenamefont {Chen}, \citenamefont {Yu},
  \citenamefont {Wang}, \citenamefont {Xia}, \citenamefont {Shi}, \citenamefont
  {Wang},\ and\ \citenamefont {Du}}]{ji_correlated_2024}%
  \BibitemOpen
  \bibfield  {author} {\bibinfo {author} {\bibfnamefont {W.}~\bibnamefont
  {Ji}}, \bibinfo {author} {\bibfnamefont {Z.}~\bibnamefont {Liu}}, \bibinfo
  {author} {\bibfnamefont {Y.}~\bibnamefont {Guo}}, \bibinfo {author}
  {\bibfnamefont {Z.}~\bibnamefont {Hu}}, \bibinfo {author} {\bibfnamefont
  {J.}~\bibnamefont {Zhou}}, \bibinfo {author} {\bibfnamefont {S.}~\bibnamefont
  {Dai}}, \bibinfo {author} {\bibfnamefont {Y.}~\bibnamefont {Chen}}, \bibinfo
  {author} {\bibfnamefont {P.}~\bibnamefont {Yu}}, \bibinfo {author}
  {\bibfnamefont {M.}~\bibnamefont {Wang}}, \bibinfo {author} {\bibfnamefont
  {K.}~\bibnamefont {Xia}}, \bibinfo {author} {\bibfnamefont {F.}~\bibnamefont
  {Shi}}, \bibinfo {author} {\bibfnamefont {Y.}~\bibnamefont {Wang}},\ and\
  \bibinfo {author} {\bibfnamefont {J.}~\bibnamefont {Du}},\ }\bibfield
  {title} {\bibinfo {title} {Correlated sensing with a solid-state quantum
  multisensor system for atomic-scale structural analysis},\ }\href
  {https://doi.org/10.1038/s41566-023-01352-4} {\bibfield  {journal} {\bibinfo
  {journal} {Nat. Photon.}\ ,\ \bibinfo {pages} {1}} (\bibinfo {year}
  {2024})}\BibitemShut
  {NoStop}%
\bibitem [{\citenamefont {Delord}\ \emph {et~al.}(2024)\citenamefont {Delord},
  \citenamefont {Monge},\ and\ \citenamefont
  {Meriles}}]{delord_correlated_2024}%
  \BibitemOpen
  \bibfield  {author} {\bibinfo {author} {\bibfnamefont {T.}~\bibnamefont
  {Delord}}, \bibinfo {author} {\bibfnamefont {R.}~\bibnamefont {Monge}},\ and\
  \bibinfo {author} {\bibfnamefont {C.~A.},\bibinfo {title} {Correlated
  spectroscopy of electric noise with color center clusters} \bibnamefont {Meriles}},\ }\href
  {https://doi.org/10.48550/arXiv.2401.07814} {\bibinfo {note} {arXiv:2401.07814} (\bibinfo {year}
  {2024})} \BibitemShut {NoStop}%
\bibitem [{\citenamefont {Bourgeois}\ \emph {et~al.}(2015)\citenamefont
  {Bourgeois}, \citenamefont {Jarmola}, \citenamefont {Siyushev}, \citenamefont
  {Gulka}, \citenamefont {Hruby}, \citenamefont {Jelezko}, \citenamefont
  {Budker},\ and\ \citenamefont {Nesladek}}]{bourgeois_photoelectric_2015}%
  \BibitemOpen
  \bibfield  {author} {\bibinfo {author} {\bibfnamefont {E.}~\bibnamefont
  {Bourgeois}}, \bibinfo {author} {\bibfnamefont {A.}~\bibnamefont {Jarmola}},
  \bibinfo {author} {\bibfnamefont {P.}~\bibnamefont {Siyushev}}, \bibinfo
  {author} {\bibfnamefont {M.}~\bibnamefont {Gulka}}, \bibinfo {author}
  {\bibfnamefont {J.}~\bibnamefont {Hruby}}, \bibinfo {author} {\bibfnamefont
  {F.}~\bibnamefont {Jelezko}}, \bibinfo {author} {\bibfnamefont
  {D.}~\bibnamefont {Budker}},\ and\ \bibinfo {author} {\bibfnamefont
  {M.}~\bibnamefont {Nesladek}},\ }\bibfield  {title} {\bibinfo {title}
  {Photoelectric detection of electron spin resonance of nitrogen-vacancy
  centres in diamond},\ }\href {https://doi.org/10.1038/ncomms9577} {\bibfield
  {journal} {\bibinfo  {journal} {Nat. Commun.}\ }\textbf {\bibinfo {volume}
  {6}},\ \bibinfo {pages} {8577} (\bibinfo {year} {2015})}\BibitemShut {NoStop}%
\bibitem [{\citenamefont {Siyushev}\ \emph {et~al.}(2019)\citenamefont
  {Siyushev}, \citenamefont {Nesladek}, \citenamefont {Bourgeois},
  \citenamefont {Gulka}, \citenamefont {Hruby}, \citenamefont {Yamamoto},
  \citenamefont {Trupke}, \citenamefont {Teraji}, \citenamefont {Isoya},\ and\
  \citenamefont {Jelezko}}]{siyushev_photoelectrical_2019}%
  \BibitemOpen
  \bibfield  {author} {\bibinfo {author} {\bibfnamefont {P.}~\bibnamefont
  {Siyushev}}, \bibinfo {author} {\bibfnamefont {M.}~\bibnamefont {Nesladek}},
  \bibinfo {author} {\bibfnamefont {E.}~\bibnamefont {Bourgeois}}, \bibinfo
  {author} {\bibfnamefont {M.}~\bibnamefont {Gulka}}, \bibinfo {author}
  {\bibfnamefont {J.}~\bibnamefont {Hruby}}, \bibinfo {author} {\bibfnamefont
  {T.}~\bibnamefont {Yamamoto}}, \bibinfo {author} {\bibfnamefont
  {M.}~\bibnamefont {Trupke}}, \bibinfo {author} {\bibfnamefont
  {T.}~\bibnamefont {Teraji}}, \bibinfo {author} {\bibfnamefont
  {J.}~\bibnamefont {Isoya}},\ and\ \bibinfo {author} {\bibfnamefont
  {F.}~\bibnamefont {Jelezko}},\ }\bibfield  {title} {\bibinfo {title}
  {Photoelectrical imaging and coherent spin-state readout of single
  nitrogen-vacancy centers in diamond},\ }\href
  {https://doi.org/10.1126/science.aav2789} {\bibfield  {journal} {\bibinfo
  {journal} {Science}\ }\textbf {\bibinfo {volume} {363}},\ \bibinfo {pages}
  {728} (\bibinfo {year} {2019})}\BibitemShut {NoStop}%
\bibitem [{\citenamefont {Bourgeois}\ \emph {et~al.}(2020)\citenamefont
  {Bourgeois}, \citenamefont {Gulka},\ and\ \citenamefont
  {Nesladek}}]{bourgeois_photoelectric_2020}%
  \BibitemOpen
  \bibfield  {author} {\bibinfo {author} {\bibfnamefont {E.}~\bibnamefont
  {Bourgeois}}, \bibinfo {author} {\bibfnamefont {M.}~\bibnamefont {Gulka}},\
  and\ \bibinfo {author} {\bibfnamefont {M.}~\bibnamefont {Nesladek}},\
  }\bibfield  {title} {\bibinfo {title} {Photoelectric {Detection} and
  {Quantum} {Readout} of {Nitrogen}-{Vacancy} {Center} {Spin} {States} in
  {Diamond}},\ }\href {https://doi.org/10.1002/adom.201902132} {\bibfield
  {journal} {\bibinfo  {journal} {Adv. Opt. Mat.}\ }\textbf
  {\bibinfo {volume} {8}},\ \bibinfo {pages} {1902132} (\bibinfo {year}
  {2020})}\BibitemShut
  {NoStop}%
\bibitem [{\citenamefont {Hruby}\ \emph {et~al.}(2022)\citenamefont {Hruby},
  \citenamefont {Gulka}, \citenamefont {Mongillo}, \citenamefont {Radu},
  \citenamefont {Petrov}, \citenamefont {Bourgeois},\ and\ \citenamefont
  {Nesladek}}]{hruby_magnetic_2022}%
  \BibitemOpen
  \bibfield  {author} {\bibinfo {author} {\bibfnamefont {J.}~\bibnamefont
  {Hruby}}, \bibinfo {author} {\bibfnamefont {M.}~\bibnamefont {Gulka}},
  \bibinfo {author} {\bibfnamefont {M.}~\bibnamefont {Mongillo}}, \bibinfo
  {author} {\bibfnamefont {I.~P.}\ \bibnamefont {Radu}}, \bibinfo {author}
  {\bibfnamefont {M.~V.}\ \bibnamefont {Petrov}}, \bibinfo {author}
  {\bibfnamefont {E.}~\bibnamefont {Bourgeois}},\ and\ \bibinfo {author}
  {\bibfnamefont {M.}~\bibnamefont {Nesladek}},\ }\bibfield  {title} {\bibinfo
  {title} {Magnetic field sensitivity of the photoelectrically read
  nitrogen-vacancy centers in diamond},\ }\href
  {https://doi.org/10.1063/5.0079667} {\bibfield  {journal} {\bibinfo
  {journal} {Appl. Phys. Lett.}\ }\textbf {\bibinfo {volume} {120}},\
  \bibinfo {pages} {162402} (\bibinfo {year} {2022})}\BibitemShut {NoStop}%
\bibitem [{\citenamefont {Bourgeois}\ \emph {et~al.}(2022)\citenamefont
  {Bourgeois}, \citenamefont {Soucek}, \citenamefont {Hruby}, \citenamefont
  {Gulka},\ and\ \citenamefont {Nesladek}}]{bourgeois_photoelectric_2022}%
  \BibitemOpen
  \bibfield  {author} {\bibinfo {author} {\bibfnamefont {E.}~\bibnamefont
  {Bourgeois}}, \bibinfo {author} {\bibfnamefont {J.}~\bibnamefont {Soucek}},
  \bibinfo {author} {\bibfnamefont {J.}~\bibnamefont {Hruby}}, \bibinfo
  {author} {\bibfnamefont {M.}~\bibnamefont {Gulka}},\ and\ \bibinfo {author}
  {\bibfnamefont {M.}~\bibnamefont {Nesladek}},\ }\bibfield  {title} {\bibinfo
  {title} {Photoelectric {Detection} of {Nitrogen}-{Vacancy} {Centers}
  {Magnetic} {Resonances} in {Diamond}: {Role} of {Charge} {Exchanges} with
  {Other} {Optoelectrically} {Active} {Defects}},\ }\href
  {https://doi.org/10.1002/qute.202100153} {\bibfield  {journal} {\bibinfo
  {journal} {Adv. Quantum Technol.}\ }\textbf {\bibinfo {volume} {5}},\
  \bibinfo {pages} {2100153} (\bibinfo {year} {2022})}\BibitemShut
  {NoStop}%
\bibitem [{\citenamefont {Todenhagen}\ and\ \citenamefont
  {Brandt}(2023)}]{todenhagen_wavelength_2023}%
  \BibitemOpen
  \bibfield  {author} {\bibinfo {author} {\bibfnamefont {L.~M.}\ \bibnamefont
  {Todenhagen}}\ and\ \bibinfo {author} {\bibfnamefont {M.~S.}\ \bibnamefont
  {Brandt}},} \bibinfo {title} {Wavelength {Dependence} of the {Electrical} and {Optical} {Readout}
  of {NV} {Centers} in {Diamond}, }\href {https://doi.org/10.48550/arXiv.2307.11830} {\bibinfo {note}
  {arXiv:2307.11830} (\bibinfo {year} {2023})} \BibitemShut {NoStop}%
\bibitem [{\citenamefont {Lozovoi}\ \emph
  {et~al.}(2020{\natexlab{b}})\citenamefont {Lozovoi}, \citenamefont
  {Jayakumar}, \citenamefont {Daw}, \citenamefont {Lakra},\ and\ \citenamefont
  {Meriles}}]{lozovoi_probing_2020}%
  \BibitemOpen
  \bibfield  {author} {\bibinfo {author} {\bibfnamefont {A.}~\bibnamefont
  {Lozovoi}}, \bibinfo {author} {\bibfnamefont {H.}~\bibnamefont {Jayakumar}},
  \bibinfo {author} {\bibfnamefont {D.}~\bibnamefont {Daw}}, \bibinfo {author}
  {\bibfnamefont {A.}~\bibnamefont {Lakra}},\ and\ \bibinfo {author}
  {\bibfnamefont {C.~A.}\ \bibnamefont {Meriles}},\ }\bibfield  {title}
  {\bibinfo {title} {Probing {Metastable} {Space}-{Charge} {Potentials} in a
  {Wide} {Band} {Gap} {Semiconductor}},\ }\href
  {https://doi.org/10.1103/PhysRevLett.125.256602} {\bibfield  {journal}
  {\bibinfo  {journal} {Phys. Rev. Lett.}\ }\textbf {\bibinfo {volume} {125}},\
  \bibinfo {pages} {256602} (\bibinfo {year} {2020}{\natexlab{b}})}\BibitemShut {NoStop}%
\bibitem [{\citenamefont {Gardill}\ \emph {et~al.}(2021)\citenamefont
  {Gardill}, \citenamefont {Kemeny}, \citenamefont {Cambria}, \citenamefont
  {Li}, \citenamefont {Dinani}, \citenamefont {Norambuena}, \citenamefont
  {Maze}, \citenamefont {Lordi},\ and\ \citenamefont
  {Kolkowitz}}]{gardill_probing_2021}%
  \BibitemOpen
  \bibfield  {author} {\bibinfo {author} {\bibfnamefont {A.}~\bibnamefont
  {Gardill}}, \bibinfo {author} {\bibfnamefont {I.}~\bibnamefont {Kemeny}},
  \bibinfo {author} {\bibfnamefont {M.~C.}\ \bibnamefont {Cambria}}, \bibinfo
  {author} {\bibfnamefont {Y.}~\bibnamefont {Li}}, \bibinfo {author}
  {\bibfnamefont {H.~T.}\ \bibnamefont {Dinani}}, \bibinfo {author}
  {\bibfnamefont {A.}~\bibnamefont {Norambuena}}, \bibinfo {author}
  {\bibfnamefont {J.~R.}\ \bibnamefont {Maze}}, \bibinfo {author}
  {\bibfnamefont {V.}~\bibnamefont {Lordi}},\ and\ \bibinfo {author}
  {\bibfnamefont {S.}~\bibnamefont {Kolkowitz}},\ }\bibfield  {title} {\bibinfo
  {title} {Probing {Charge} {Dynamics} in {Diamond} with an {Individual}
  {Color} {Center}},\ }\href {https://doi.org/10.1021/acs.nanolett.1c02250}
  {\bibfield  {journal} {\bibinfo  {journal} {Nano Lett.}\ }\textbf {\bibinfo
  {volume} {21}},\ \bibinfo {pages} {6960} (\bibinfo {year} {2021})}\BibitemShut {NoStop}%
\bibitem [{\citenamefont {Lozovoi}\ \emph {et~al.}(2022)\citenamefont
  {Lozovoi}, \citenamefont {Vizkelethy}, \citenamefont {Bielejec},\ and\
  \citenamefont {Meriles}}]{lozovoi_imaging_2022}%
  \BibitemOpen
  \bibfield  {author} {\bibinfo {author} {\bibfnamefont {A.}~\bibnamefont
  {Lozovoi}}, \bibinfo {author} {\bibfnamefont {G.}~\bibnamefont {Vizkelethy}},
  \bibinfo {author} {\bibfnamefont {E.}~\bibnamefont {Bielejec}},\ and\
  \bibinfo {author} {\bibfnamefont {C.~A.}\ \bibnamefont {Meriles}},\
  }\bibfield  {title} {\bibinfo {title} {Imaging dark charge emitters in
  diamond via carrier-to-photon conversion},\ }\href
  {https://doi.org/10.1126/sciadv.abl9402} {\bibfield  {journal} {\bibinfo
  {journal} {Science Advances}\ }\textbf {\bibinfo {volume} {8}},\ \bibinfo
  {pages} {eabl9402} (\bibinfo {year} {2022})}\BibitemShut {NoStop}%
\bibitem [{\citenamefont {Lozovoi}\ \emph {et~al.}(2021)\citenamefont
  {Lozovoi}, \citenamefont {Jayakumar}, \citenamefont {Daw}, \citenamefont
  {Vizkelethy}, \citenamefont {Bielejec}, \citenamefont {Doherty},
  \citenamefont {Flick},\ and\ \citenamefont {Meriles}}]{lozovoi_optical_2021}%
  \BibitemOpen
  \bibfield  {author} {\bibinfo {author} {\bibfnamefont {A.}~\bibnamefont
  {Lozovoi}}, \bibinfo {author} {\bibfnamefont {H.}~\bibnamefont {Jayakumar}},
  \bibinfo {author} {\bibfnamefont {D.}~\bibnamefont {Daw}}, \bibinfo {author}
  {\bibfnamefont {G.}~\bibnamefont {Vizkelethy}}, \bibinfo {author}
  {\bibfnamefont {E.}~\bibnamefont {Bielejec}}, \bibinfo {author}
  {\bibfnamefont {M.~W.}\ \bibnamefont {Doherty}}, \bibinfo {author}
  {\bibfnamefont {J.}~\bibnamefont {Flick}},\ and\ \bibinfo {author}
  {\bibfnamefont {C.~A.}\ \bibnamefont {Meriles}},\ }\bibfield  {title}
  {\bibinfo {title} {Optical activation and detection of charge transport
  between individual colour centres in diamond},\ }\href
  {https://doi.org/10.1038/s41928-021-00656-z} {\bibfield  {journal} {\bibinfo
  {journal} {Nat Electron}\ }\textbf {\bibinfo {volume} {4}},\ \bibinfo {pages}
  {717} (\bibinfo {year} {2021})}\BibitemShut {NoStop}%
\bibitem [{\citenamefont {Lozovoi}\ \emph {et~al.}(2023)\citenamefont
  {Lozovoi}, \citenamefont {Chen}, \citenamefont {Vizkelethy}, \citenamefont
  {Bielejec}, \citenamefont {Flick}, \citenamefont {Doherty},\ and\
  \citenamefont {Meriles}}]{lozovoi_detection_2023}%
  \BibitemOpen
  \bibfield  {author} {\bibinfo {author} {\bibfnamefont {A.}~\bibnamefont
  {Lozovoi}}, \bibinfo {author} {\bibfnamefont {Y.}~\bibnamefont {Chen}},
  \bibinfo {author} {\bibfnamefont {G.}~\bibnamefont {Vizkelethy}}, \bibinfo
  {author} {\bibfnamefont {E.}~\bibnamefont {Bielejec}}, \bibinfo {author}
  {\bibfnamefont {J.}~\bibnamefont {Flick}}, \bibinfo {author} {\bibfnamefont
  {M.~W.}\ \bibnamefont {Doherty}},\ and\ \bibinfo {author} {\bibfnamefont
  {C.~A.}\ \bibnamefont {Meriles}},\ }\bibfield  {title} {\bibinfo {title}
  {Detection and {Modeling} of {Hole} {Capture} by {Single} {Point} {Defects}
  under {Variable} {Electric} {Fields}},\ }\href
  {https://doi.org/10.1021/acs.nanolett.3c00860} {\bibfield  {journal}
  {\bibinfo  {journal} {Nano Lett.}\ }\textbf {\bibinfo {volume} {23}},\
  \bibinfo {pages} {4495} (\bibinfo {year} {2023})}\BibitemShut {NoStop}%
\bibitem [{\citenamefont {Zheng}\ \emph {et~al.}(2022)\citenamefont {Zheng},
  \citenamefont {Bian}, \citenamefont {Chen}, \citenamefont {Shen},
  \citenamefont {Zhang}, \citenamefont {Stöhr}, \citenamefont {Denisenko},
  \citenamefont {Wrachtrup}, \citenamefont {Yang},\ and\ \citenamefont
  {Jiang}}]{zheng_coherence_2022}%
  \BibitemOpen
  \bibfield  {author} {\bibinfo {author} {\bibfnamefont {W.}~\bibnamefont
  {Zheng}}, \bibinfo {author} {\bibfnamefont {K.}~\bibnamefont {Bian}},
  \bibinfo {author} {\bibfnamefont {X.}~\bibnamefont {Chen}}, \bibinfo {author}
  {\bibfnamefont {Y.}~\bibnamefont {Shen}}, \bibinfo {author} {\bibfnamefont
  {S.}~\bibnamefont {Zhang}}, \bibinfo {author} {\bibfnamefont
  {R.}~\bibnamefont {Stöhr}}, \bibinfo {author} {\bibfnamefont
  {A.}~\bibnamefont {Denisenko}}, \bibinfo {author} {\bibfnamefont
  {J.}~\bibnamefont {Wrachtrup}}, \bibinfo {author} {\bibfnamefont
  {S.}~\bibnamefont {Yang}},\ and\ \bibinfo {author} {\bibfnamefont
  {Y.}~\bibnamefont {Jiang}},\ }\bibfield  {title} {\bibinfo {title} {Coherence
  enhancement of solid-state qubits by local manipulation of the electron spin
  bath},\ }\href {https://doi.org/10.1038/s41567-022-01719-4} {\bibfield
  {journal} {\bibinfo  {journal} {Nat. Phys.}\ }\textbf {\bibinfo {volume}
  {18}},\ \bibinfo {pages} {1317} (\bibinfo {year} {2022})}\BibitemShut {NoStop}%
\bibitem [{\citenamefont {Görlitz}\ \emph {et~al.}(2022)\citenamefont
  {Görlitz}, \citenamefont {Herrmann}, \citenamefont {Fuchs}, \citenamefont
  {Iwasaki}, \citenamefont {Taniguchi}, \citenamefont {Rogalla}, \citenamefont
  {Hardeman}, \citenamefont {Colard}, \citenamefont {Markham}, \citenamefont
  {Hatano},\ and\ \citenamefont {Becher}}]{gorlitz_coherence_2022}%
  \BibitemOpen
  \bibfield  {author} {\bibinfo {author} {\bibfnamefont {J.}~\bibnamefont
  {Görlitz}}, \bibinfo {author} {\bibfnamefont {D.}~\bibnamefont {Herrmann}},
  \bibinfo {author} {\bibfnamefont {P.}~\bibnamefont {Fuchs}}, \bibinfo
  {author} {\bibfnamefont {T.}~\bibnamefont {Iwasaki}}, \bibinfo {author}
  {\bibfnamefont {T.}~\bibnamefont {Taniguchi}}, \bibinfo {author}
  {\bibfnamefont {D.}~\bibnamefont {Rogalla}}, \bibinfo {author} {\bibfnamefont
  {D.}~\bibnamefont {Hardeman}}, \bibinfo {author} {\bibfnamefont {P.-O.}\
  \bibnamefont {Colard}}, \bibinfo {author} {\bibfnamefont {M.}~\bibnamefont
  {Markham}}, \bibinfo {author} {\bibfnamefont {M.}~\bibnamefont {Hatano}},\
  and\ \bibinfo {author} {\bibfnamefont {C.}~\bibnamefont {Becher}},\
  }\bibfield  {title} {\bibinfo {title} {Coherence of a charge stabilised
  tin-vacancy spin in diamond},\ }\href
  {https://doi.org/10.1038/s41534-022-00552-0} {\bibfield  {journal} {\bibinfo
  {journal} {npj Quantum Inf}\ }\textbf {\bibinfo {volume} {8}},\ \bibinfo
  {pages} {1} (\bibinfo {year} {2022})}\BibitemShut {NoStop}%
\bibitem [{\citenamefont {Zuber}\ \emph {et~al.}(2023)\citenamefont {Zuber},
  \citenamefont {Li}, \citenamefont {Puigibert}, \citenamefont {Happacher},
  \citenamefont {Reiser}, \citenamefont {Shields},\ and\ \citenamefont
  {Maletinsky}}]{zuber_shallow_2023}%
  \BibitemOpen
  \bibfield  {author} {\bibinfo {author} {\bibfnamefont {J.~A.}\ \bibnamefont
  {Zuber}}, \bibinfo {author} {\bibfnamefont {M.}~\bibnamefont {Li}}, \bibinfo
  {author} {\bibfnamefont {M.~l.~G.}\ \bibnamefont {Puigibert}}, \bibinfo
  {author} {\bibfnamefont {J.}~\bibnamefont {Happacher}}, \bibinfo {author}
  {\bibfnamefont {P.}~\bibnamefont {Reiser}}, \bibinfo {author} {\bibfnamefont
  {B.~J.}\ \bibnamefont {Shields}},\ and\ \bibinfo {author} {\bibfnamefont
  {P.}~\bibnamefont {Maletinsky}},\ }\bibinfo {title} {Shallow
  {Silicon} {Vacancy} {Centers} with lifetime-limited optical linewidths in
  {Diamond} {Nanostructures}, }\href
  {https://doi.org/10.48550/arXiv.2307.12753} {\bibinfo {note}
  {arXiv:2307.12753} (\bibinfo {year} {2023})} \BibitemShut {NoStop}%
\bibitem [{\citenamefont {Edmonds}\ \emph {et~al.}(2012)\citenamefont
  {Edmonds}, \citenamefont {D’Haenens-Johansson}, \citenamefont {Cruddace},
  \citenamefont {Newton}, \citenamefont {Fu}, \citenamefont {Santori},
  \citenamefont {Beausoleil}, \citenamefont {Twitchen},\ and\ \citenamefont
  {Markham}}]{edmonds_production_2012}%
  \BibitemOpen
  \bibfield  {author} {\bibinfo {author} {\bibfnamefont {A.~M.}\ \bibnamefont
  {Edmonds}}, \bibinfo {author} {\bibfnamefont {U.~F.~S.}\ \bibnamefont
  {D’Haenens-Johansson}}, \bibinfo {author} {\bibfnamefont {R.~J.}\
  \bibnamefont {Cruddace}}, \bibinfo {author} {\bibfnamefont {M.~E.}\
  \bibnamefont {Newton}}, \bibinfo {author} {\bibfnamefont {K.-M.~C.}\
  \bibnamefont {Fu}}, \bibinfo {author} {\bibfnamefont {C.}~\bibnamefont
  {Santori}}, \bibinfo {author} {\bibfnamefont {R.~G.}\ \bibnamefont
  {Beausoleil}}, \bibinfo {author} {\bibfnamefont {D.~J.}\ \bibnamefont
  {Twitchen}},\ and\ \bibinfo {author} {\bibfnamefont {M.~L.}\ \bibnamefont
  {Markham}},\ }\bibfield  {title} {\bibinfo {title} {Production of oriented
  nitrogen-vacancy color centers in synthetic diamond},\ }\href
  {https://doi.org/10.1103/PhysRevB.86.035201} {\bibfield  {journal} {\bibinfo
  {journal} {Phys. Rev. B}\ }\textbf {\bibinfo {volume} {86}},\ \bibinfo
  {pages} {035201} (\bibinfo {year} {2012})}\BibitemShut {NoStop}%
\bibitem [{\citenamefont {Dhomkar}\ \emph {et~al.}(2018)\citenamefont
  {Dhomkar}, \citenamefont {Zangara}, \citenamefont {Henshaw},\ and\
  \citenamefont {Meriles}}]{dhomkar_-demand_2018}%
  \BibitemOpen
  \bibfield  {author} {\bibinfo {author} {\bibfnamefont {S.}~\bibnamefont
  {Dhomkar}}, \bibinfo {author} {\bibfnamefont {P.~R.}\ \bibnamefont
  {Zangara}}, \bibinfo {author} {\bibfnamefont {J.}~\bibnamefont {Henshaw}},\
  and\ \bibinfo {author} {\bibfnamefont {C.~A.}\ \bibnamefont {Meriles}},\
  }\bibfield  {title} {\bibinfo {title} {On-{Demand} {Generation} of {Neutral}
  and {Negatively} {Charged} {Silicon}-{Vacancy} {Centers} in {Diamond}},\
  }\href {https://doi.org/10.1103/PhysRevLett.120.117401} {\bibfield  {journal}
  {\bibinfo  {journal} {Phys. Rev. Lett.}\ }\textbf {\bibinfo {volume} {120}},\
  \bibinfo {pages} {117401} (\bibinfo {year} {2018})}\BibitemShut {NoStop}%
\bibitem [{\citenamefont {Gali}\ and\ \citenamefont
  {Maze}(2013)}]{gali_ab_2013}%
  \BibitemOpen
  \bibfield  {author} {\bibinfo {author} {\bibfnamefont {A.}~\bibnamefont
  {Gali}}\ and\ \bibinfo {author} {\bibfnamefont {J.~R.}\ \bibnamefont
  {Maze}},\ }\bibfield  {title} {\bibinfo {title} {Ab initio study of the split
  silicon-vacancy defect in diamond: {Electronic} structure and related
  properties},\ }\href {https://doi.org/10.1103/PhysRevB.88.235205} {\bibfield
  {journal} {\bibinfo  {journal} {Phys. Rev. B}\ }\textbf {\bibinfo {volume}
  {88}},\ \bibinfo {pages} {235205} (\bibinfo {year} {2013})}\BibitemShut {NoStop}%
\bibitem [{\citenamefont {Thiering}\ and\ \citenamefont
  {Gali}(2018)}]{thiering_ab_2018}%
  \BibitemOpen
  \bibfield  {author} {\bibinfo {author} {\bibfnamefont {G.}~\bibnamefont
  {Thiering}}\ and\ \bibinfo {author} {\bibfnamefont {A.}~\bibnamefont
  {Gali}},\ }\bibfield  {title} {\bibinfo {title} {Ab {Initio}
  {Magneto}-{Optical} {Spectrum} of {Group}-{IV} {Vacancy} {Color} {Centers} in
  {Diamond}},\ }\href {https://doi.org/10.1103/PhysRevX.8.021063} {\bibfield
  {journal} {\bibinfo  {journal} {Phys. Rev. X}\ }\textbf {\bibinfo {volume}
  {8}},\ \bibinfo {pages} {021063} (\bibinfo {year} {2018})}\BibitemShut {NoStop}%
\bibitem [{\citenamefont {Jayakumar}\ \emph {et~al.}(2016)\citenamefont
  {Jayakumar}, \citenamefont {Henshaw}, \citenamefont {Dhomkar}, \citenamefont
  {Pagliero}, \citenamefont {Laraoui}, \citenamefont {Manson}, \citenamefont
  {Albu}, \citenamefont {Doherty},\ and\ \citenamefont
  {Meriles}}]{jayakumar_optical_2016}%
  \BibitemOpen
  \bibfield  {author} {\bibinfo {author} {\bibfnamefont {H.}~\bibnamefont
  {Jayakumar}}, \bibinfo {author} {\bibfnamefont {J.}~\bibnamefont {Henshaw}},
  \bibinfo {author} {\bibfnamefont {S.}~\bibnamefont {Dhomkar}}, \bibinfo
  {author} {\bibfnamefont {D.}~\bibnamefont {Pagliero}}, \bibinfo {author}
  {\bibfnamefont {A.}~\bibnamefont {Laraoui}}, \bibinfo {author} {\bibfnamefont
  {N.~B.}\ \bibnamefont {Manson}}, \bibinfo {author} {\bibfnamefont
  {R.}~\bibnamefont {Albu}}, \bibinfo {author} {\bibfnamefont {M.~W.}\
  \bibnamefont {Doherty}},\ and\ \bibinfo {author} {\bibfnamefont {C.~A.}\
  \bibnamefont {Meriles}},\ }\bibfield  {title} {\bibinfo {title} {Optical
  patterning of trapped charge in nitrogen-doped diamond},\ }\href
  {https://doi.org/10.1038/ncomms12660} {\bibfield  {journal} {\bibinfo
  {journal} {Nat Commun}\ }\textbf {\bibinfo {volume} {7}},\ \bibinfo {pages}
  {12660} (\bibinfo {year} {2016})}\BibitemShut {NoStop}%
\bibitem [{\citenamefont {Jayakumar}\ \emph {et~al.}(2020)\citenamefont
  {Jayakumar}, \citenamefont {Lozovoi}, \citenamefont {Daw},\ and\
  \citenamefont {Meriles}}]{jayakumar_long-term_2020}%
  \BibitemOpen
  \bibfield  {author} {\bibinfo {author} {\bibfnamefont {H.}~\bibnamefont
  {Jayakumar}}, \bibinfo {author} {\bibfnamefont {A.}~\bibnamefont {Lozovoi}},
  \bibinfo {author} {\bibfnamefont {D.}~\bibnamefont {Daw}},\ and\ \bibinfo
  {author} {\bibfnamefont {C.}~\bibnamefont {Meriles}},\ }\bibfield  {title}
  {\bibinfo {title} {Long-{Term} {Spin} {State} {Storage} {Using} {Ancilla}
  {Charge} {Memories}},\ }\href
  {https://doi.org/10.1103/PhysRevLett.125.236601} {\bibfield  {journal}
  {\bibinfo  {journal} {Phys. Rev. Lett.}\ }\textbf {\bibinfo {volume} {125}},\
  \bibinfo {pages} {236601} (\bibinfo {year} {2020})}\BibitemShut {NoStop}%
\bibitem [{\citenamefont {Bourgeois}\ \emph {et~al.}(2017)\citenamefont
  {Bourgeois}, \citenamefont {Londero}, \citenamefont {Buczak}, \citenamefont
  {Hruby}, \citenamefont {Gulka}, \citenamefont {Balasubramaniam},
  \citenamefont {Wachter}, \citenamefont {Stursa}, \citenamefont {Dobes},
  \citenamefont {Aumayr}, \citenamefont {Trupke}, \citenamefont {Gali},\ and\
  \citenamefont {Nesladek}}]{bourgeois_enhanced_2017}%
  \BibitemOpen
  \bibfield  {author} {\bibinfo {author} {\bibfnamefont {E.}~\bibnamefont
  {Bourgeois}}, \bibinfo {author} {\bibfnamefont {E.}~\bibnamefont {Londero}},
  \bibinfo {author} {\bibfnamefont {K.}~\bibnamefont {Buczak}}, \bibinfo
  {author} {\bibfnamefont {J.}~\bibnamefont {Hruby}}, \bibinfo {author}
  {\bibfnamefont {M.}~\bibnamefont {Gulka}}, \bibinfo {author} {\bibfnamefont
  {Y.}~\bibnamefont {Balasubramaniam}}, \bibinfo {author} {\bibfnamefont
  {G.}~\bibnamefont {Wachter}}, \bibinfo {author} {\bibfnamefont
  {J.}~\bibnamefont {Stursa}}, \bibinfo {author} {\bibfnamefont
  {K.}~\bibnamefont {Dobes}}, \bibinfo {author} {\bibfnamefont
  {F.}~\bibnamefont {Aumayr}}, \bibinfo {author} {\bibfnamefont
  {M.}~\bibnamefont {Trupke}}, \bibinfo {author} {\bibfnamefont
  {A.}~\bibnamefont {Gali}},\ and\ \bibinfo {author} {\bibfnamefont
  {M.}~\bibnamefont {Nesladek}},\ }\bibfield  {title} {\bibinfo {title}
  {Enhanced photoelectric detection of {NV} magnetic resonances in diamond
  under dual-beam excitation},\ }\href
  {https://doi.org/10.1103/PhysRevB.95.041402} {\bibfield  {journal} {\bibinfo
  {journal} {Phys. Rev. B}\ }\textbf {\bibinfo {volume} {95}},\ \bibinfo
  {pages} {041402} (\bibinfo {year} {2017})}\BibitemShut {NoStop}%
\bibitem [{\citenamefont {Manson}\ \emph {et~al.}(2013)\citenamefont {Manson},
  \citenamefont {Beha}, \citenamefont {Batalov}, \citenamefont {Rogers},
  \citenamefont {Doherty}, \citenamefont {Bratschitsch},\ and\ \citenamefont
  {Leitenstorfer}}]{manson_assignment_2013}%
  \BibitemOpen
  \bibfield  {author} {\bibinfo {author} {\bibfnamefont {N.~B.}\ \bibnamefont
  {Manson}}, \bibinfo {author} {\bibfnamefont {K.}~\bibnamefont {Beha}},
  \bibinfo {author} {\bibfnamefont {A.}~\bibnamefont {Batalov}}, \bibinfo
  {author} {\bibfnamefont {L.~J.}\ \bibnamefont {Rogers}}, \bibinfo {author}
  {\bibfnamefont {M.~W.}\ \bibnamefont {Doherty}}, \bibinfo {author}
  {\bibfnamefont {R.}~\bibnamefont {Bratschitsch}},\ and\ \bibinfo {author}
  {\bibfnamefont {A.}~\bibnamefont {Leitenstorfer}},\ }\bibfield  {title}
  {\bibinfo {title} {Assignment of the {NV}$^0$ 575-nm zero-phonon line in diamond to a
  $^2E$-$^2A_2$ transition},\ }\href {https://doi.org/10.1103/PhysRevB.87.155209} {\bibfield
  {journal} {\bibinfo  {journal} {Phys. Rev. B}\ }\textbf {\bibinfo {volume}
  {87}},\ \bibinfo {pages} {155209} (\bibinfo {year} {2013})}\BibitemShut {NoStop}%
\bibitem [{\citenamefont {Meirzada}\ \emph {et~al.}(2018)\citenamefont
  {Meirzada}, \citenamefont {Hovav}, \citenamefont {Wolf},\ and\ \citenamefont
  {Bar-Gill}}]{meirzada_negative_2018}%
  \BibitemOpen
  \bibfield  {author} {\bibinfo {author} {\bibfnamefont {I.}~\bibnamefont
  {Meirzada}}, \bibinfo {author} {\bibfnamefont {Y.}~\bibnamefont {Hovav}},
  \bibinfo {author} {\bibfnamefont {S.~A.}\ \bibnamefont {Wolf}},\ and\
  \bibinfo {author} {\bibfnamefont {N.}~\bibnamefont {Bar-Gill}},\ }\bibfield
  {title} {\bibinfo {title} {Negative charge enhancement of near-surface
  nitrogen vacancy centers by multicolor excitation},\ }\href
  {https://doi.org/10.1103/PhysRevB.98.245411} {\bibfield  {journal} {\bibinfo
  {journal} {Phys. Rev. B}\ }\textbf {\bibinfo {volume} {98}},\ \bibinfo
  {pages} {245411} (\bibinfo {year} {2018})}\BibitemShut {NoStop}%
\bibitem [{\citenamefont {Nesládek}\ \emph {et~al.}(1998)\citenamefont
  {Nesládek}, \citenamefont {Stals}, \citenamefont {Stesmans}, \citenamefont
  {Iakoubovskij}, \citenamefont {Adriaenssens}, \citenamefont {Rosa},\ and\
  \citenamefont {Vaněček}}]{nesladek_dominant_1998}%
  \BibitemOpen
  \bibfield  {author} {\bibinfo {author} {\bibfnamefont {M.}~\bibnamefont
  {Nesládek}}, \bibinfo {author} {\bibfnamefont {L.~M.}\ \bibnamefont
  {Stals}}, \bibinfo {author} {\bibfnamefont {A.}~\bibnamefont {Stesmans}},
  \bibinfo {author} {\bibfnamefont {K.}~\bibnamefont {Iakoubovskij}}, \bibinfo
  {author} {\bibfnamefont {G.~J.}\ \bibnamefont {Adriaenssens}}, \bibinfo
  {author} {\bibfnamefont {J.}~\bibnamefont {Rosa}},\ and\ \bibinfo {author}
  {\bibfnamefont {M.}~\bibnamefont {Vaněček}},\ }\bibfield  {title} {\bibinfo
  {title} {Dominant defect levels in diamond thin films: {A} photocurrent and
  electron paramagnetic resonance study},\ }\href
  {https://doi.org/10.1063/1.121632} {\bibfield  {journal} {\bibinfo  {journal}
  {Applied Physics Letters}\ }\textbf {\bibinfo {volume} {72}},\ \bibinfo
  {pages} {3306} (\bibinfo {year} {1998})}\BibitemShut {NoStop}%
\bibitem [{\citenamefont {Rosa}\ \emph {et~al.}(1999)\citenamefont {Rosa},
  \citenamefont {Vaněček}, \citenamefont {Nesládek},\ and\ \citenamefont
  {Stals}}]{rosa_photoionization_1999}%
  \BibitemOpen
  \bibfield  {author} {\bibinfo {author} {\bibfnamefont {J.}~\bibnamefont
  {Rosa}}, \bibinfo {author} {\bibfnamefont {M.}~\bibnamefont {Vaněček}},
  \bibinfo {author} {\bibfnamefont {M.}~\bibnamefont {Nesládek}},\ and\
  \bibinfo {author} {\bibfnamefont {L.~M.}\ \bibnamefont {Stals}},\ }\bibfield
  {title} {\bibinfo {title} {Photoionization cross-section of dominant defects
  in {CVD} diamond},\ }\href {https://doi.org/10.1016/S0925-9635(98)00354-9}
  {\bibfield  {journal} {\bibinfo  {journal} {Diamond and Related Materials}\
  }\textbf {\bibinfo {volume} {8}},\ \bibinfo {pages} {721} (\bibinfo {year}
  {1999})}\BibitemShut {NoStop}%
\bibitem [{\citenamefont {Jones}\ \emph {et~al.}(2009)\citenamefont {Jones},
  \citenamefont {Goss},\ and\ \citenamefont {Briddon}}]{jones_acceptor_2009}%
  \BibitemOpen
  \bibfield  {author} {\bibinfo {author} {\bibfnamefont {R.}~\bibnamefont
  {Jones}}, \bibinfo {author} {\bibfnamefont {J.~P.}\ \bibnamefont {Goss}},\
  and\ \bibinfo {author} {\bibfnamefont {P.~R.}\ \bibnamefont {Briddon}},\
  }\bibfield  {title} {\bibinfo {title} {Acceptor level of nitrogen in diamond
  and the 270-nm absorption band},\ }\href
  {https://doi.org/10.1103/PhysRevB.80.033205} {\bibfield  {journal} {\bibinfo
  {journal} {Phys. Rev. B}\ }\textbf {\bibinfo {volume} {80}},\ \bibinfo
  {pages} {033205} (\bibinfo {year} {2009})}\BibitemShut {NoStop}%
\bibitem [{\citenamefont {Razinkovas}\ \emph {et~al.}(2021)\citenamefont
  {Razinkovas}, \citenamefont {Maciaszek}, \citenamefont {Reinhard},
  \citenamefont {Doherty},\ and\ \citenamefont
  {Alkauskas}}]{razinkovas_photoionization_2021}%
  \BibitemOpen
  \bibfield  {author} {\bibinfo {author} {\bibfnamefont {L.}~\bibnamefont
  {Razinkovas}}, \bibinfo {author} {\bibfnamefont {M.}~\bibnamefont
  {Maciaszek}}, \bibinfo {author} {\bibfnamefont {F.}~\bibnamefont {Reinhard}},
  \bibinfo {author} {\bibfnamefont {M.~W.}\ \bibnamefont {Doherty}},\ and\
  \bibinfo {author} {\bibfnamefont {A.}~\bibnamefont {Alkauskas}},\ }\bibfield
  {title} {\bibinfo {title} {Photoionization of negatively charged {NV} centers
  in diamond: {Theory} and ab initio calculations},\ }\href
  {https://doi.org/10.1103/PhysRevB.104.235301} {\bibfield  {journal} {\bibinfo
   {journal} {Phys. Rev. B}\ }\textbf {\bibinfo {volume} {104}},\ \bibinfo
  {pages} {235301} (\bibinfo {year} {2021})}\BibitemShut {NoStop}%
\bibitem [{\citenamefont {Lin}\ \emph {et~al.}(2022)\citenamefont {Lin},
  \citenamefont {Wang}, \citenamefont {Li}, \citenamefont {Zhou}, \citenamefont
  {Xu}, \citenamefont {Li},\ and\ \citenamefont {Guo}}]{lin_anti-stokes_2022}%
  \BibitemOpen
  \bibfield  {author} {\bibinfo {author} {\bibfnamefont {W.-X.}\ \bibnamefont
  {Lin}}, \bibinfo {author} {\bibfnamefont {J.-F.}\ \bibnamefont {Wang}},
  \bibinfo {author} {\bibfnamefont {Q.}~\bibnamefont {Li}}, \bibinfo {author}
  {\bibfnamefont {J.-Y.}\ \bibnamefont {Zhou}}, \bibinfo {author}
  {\bibfnamefont {J.-S.}\ \bibnamefont {Xu}}, \bibinfo {author} {\bibfnamefont
  {C.-F.}\ \bibnamefont {Li}},\ and\ \bibinfo {author} {\bibfnamefont {G.-C.}\
  \bibnamefont {Guo}},\ }\bibfield  {title} {\bibinfo {title} {Anti-{Stokes}
  excitation of optically active point defects in semiconductor materials},\
  }\href {https://doi.org/10.1088/2633-4356/ac989a} {\bibfield  {journal}
  {\bibinfo  {journal} {Mater. Quantum. Technol.}\ }\textbf {\bibinfo {volume}
  {2}},\ \bibinfo {pages} {042001} (\bibinfo {year} {2022})}\BibitemShut {NoStop}%
\bibitem [{\citenamefont {Tran}\ \emph {et~al.}(2019)\citenamefont {Tran},
  \citenamefont {Regan}, \citenamefont {Ekimov}, \citenamefont {Mu},
  \citenamefont {Zhou}, \citenamefont {Gao}, \citenamefont {Narang},
  \citenamefont {Solntsev}, \citenamefont {Toth}, \citenamefont {Aharonovich},\
  and\ \citenamefont {Bradac}}]{tran_anti-stokes_2019}%
  \BibitemOpen
  \bibfield  {author} {\bibinfo {author} {\bibfnamefont {T.~T.}\ \bibnamefont
  {Tran}}, \bibinfo {author} {\bibfnamefont {B.}~\bibnamefont {Regan}},
  \bibinfo {author} {\bibfnamefont {E.~A.}\ \bibnamefont {Ekimov}}, \bibinfo
  {author} {\bibfnamefont {Z.}~\bibnamefont {Mu}}, \bibinfo {author}
  {\bibfnamefont {Y.}~\bibnamefont {Zhou}}, \bibinfo {author} {\bibfnamefont
  {W.-b.}\ \bibnamefont {Gao}}, \bibinfo {author} {\bibfnamefont
  {P.}~\bibnamefont {Narang}}, \bibinfo {author} {\bibfnamefont {A.~S.}\
  \bibnamefont {Solntsev}}, \bibinfo {author} {\bibfnamefont {M.}~\bibnamefont
  {Toth}}, \bibinfo {author} {\bibfnamefont {I.}~\bibnamefont {Aharonovich}},\
  and\ \bibinfo {author} {\bibfnamefont {C.}~\bibnamefont {Bradac}},\
  }\bibfield  {title} {\bibinfo {title} {Anti-{Stokes} excitation of
  solid-state quantum emitters for nanoscale thermometry},\ }\href
  {https://doi.org/10.1126/sciadv.aav9180} {\bibfield  {journal} {\bibinfo
  {journal} {Science Advances}\ }\textbf {\bibinfo {volume} {5}},\ \bibinfo
  {pages} {eaav9180} (\bibinfo {year} {2019})}\BibitemShut {NoStop}%
\bibitem [{\citenamefont {Gao}\ \emph {et~al.}(2018)\citenamefont {Gao},
  \citenamefont {Tan}, \citenamefont {Liu}, \citenamefont {Ren}, \citenamefont
  {Sun}, \citenamefont {Meng}, \citenamefont {Lu}, \citenamefont {Tan},
  \citenamefont {Shan},\ and\ \citenamefont
  {Zhang}}]{gao_phonon-assisted_2018}%
  \BibitemOpen
  \bibfield  {author} {\bibinfo {author} {\bibfnamefont {Y.-F.}\ \bibnamefont
  {Gao}}, \bibinfo {author} {\bibfnamefont {Q.-H.}\ \bibnamefont {Tan}},
  \bibinfo {author} {\bibfnamefont {X.-L.}\ \bibnamefont {Liu}}, \bibinfo
  {author} {\bibfnamefont {S.-L.}\ \bibnamefont {Ren}}, \bibinfo {author}
  {\bibfnamefont {Y.-J.}\ \bibnamefont {Sun}}, \bibinfo {author} {\bibfnamefont
  {D.}~\bibnamefont {Meng}}, \bibinfo {author} {\bibfnamefont {Y.-J.}\
  \bibnamefont {Lu}}, \bibinfo {author} {\bibfnamefont {P.-H.}\ \bibnamefont
  {Tan}}, \bibinfo {author} {\bibfnamefont {C.-X.}\ \bibnamefont {Shan}},\ and\
  \bibinfo {author} {\bibfnamefont {J.}~\bibnamefont {Zhang}},\ }\bibfield
  {title} {\bibinfo {title} {Phonon-{Assisted} {Photoluminescence}
  {Up}-{Conversion} of {Silicon}-{Vacancy} {Centers} in {Diamond}},\ }\href
  {https://doi.org/10.1021/acs.jpclett.8b02862} {\bibfield  {journal} {\bibinfo
   {journal} {J. Phys. Chem. Lett.}\ }\textbf {\bibinfo {volume} {9}},\
  \bibinfo {pages} {6656} (\bibinfo {year} {2018})}\BibitemShut {NoStop}%
\bibitem [{\citenamefont {Gao}\ \emph {et~al.}(2022)\citenamefont {Gao},
  \citenamefont {Lai}, \citenamefont {Sun}, \citenamefont {Liu}, \citenamefont
  {Lin}, \citenamefont {Tan}, \citenamefont {Shan},\ and\ \citenamefont
  {Zhang}}]{gao_charge_2022}%
  \BibitemOpen
  \bibfield  {author} {\bibinfo {author} {\bibfnamefont {Y.-F.}\ \bibnamefont
  {Gao}}, \bibinfo {author} {\bibfnamefont {J.-M.}\ \bibnamefont {Lai}},
  \bibinfo {author} {\bibfnamefont {Y.-J.}\ \bibnamefont {Sun}}, \bibinfo
  {author} {\bibfnamefont {X.-L.}\ \bibnamefont {Liu}}, \bibinfo {author}
  {\bibfnamefont {C.-N.}\ \bibnamefont {Lin}}, \bibinfo {author} {\bibfnamefont
  {P.-H.}\ \bibnamefont {Tan}}, \bibinfo {author} {\bibfnamefont {C.-X.}\
  \bibnamefont {Shan}},\ and\ \bibinfo {author} {\bibfnamefont
  {J.}~\bibnamefont {Zhang}},\ }\bibfield  {title} {\bibinfo {title} {Charge
  {State} {Manipulation} of {NV} {Centers} in {Diamond} under
  {Phonon}-{Assisted} {Anti}-{Stokes} {Excitation} of {NV0}},\ }\href
  {https://doi.org/10.1021/acsphotonics.1c01928} {\bibfield  {journal}
  {\bibinfo  {journal} {ACS Photonics}\ }\textbf {\bibinfo {volume} {9}},\
  \bibinfo {pages} {1605} (\bibinfo {year} {2022})}\BibitemShut {NoStop}%
\bibitem [{\citenamefont {Siyushev}\ \emph {et~al.}(2013)\citenamefont
  {Siyushev}, \citenamefont {Pinto}, \citenamefont {Vörös}, \citenamefont
  {Gali}, \citenamefont {Jelezko},\ and\ \citenamefont
  {Wrachtrup}}]{siyushev_optically_2013}%
  \BibitemOpen
  \bibfield  {author} {\bibinfo {author} {\bibfnamefont {P.}~\bibnamefont
  {Siyushev}}, \bibinfo {author} {\bibfnamefont {H.}~\bibnamefont {Pinto}},
  \bibinfo {author} {\bibfnamefont {M.}~\bibnamefont {Vörös}}, \bibinfo
  {author} {\bibfnamefont {A.}~\bibnamefont {Gali}}, \bibinfo {author}
  {\bibfnamefont {F.}~\bibnamefont {Jelezko}},\ and\ \bibinfo {author}
  {\bibfnamefont {J.}~\bibnamefont {Wrachtrup}},\ }\bibfield  {title} {\bibinfo
  {title} {Optically {Controlled} {Switching} of the {Charge} {State} of a
  {Single} {Nitrogen}-{Vacancy} {Center} in {Diamond} at {Cryogenic}
  {Temperatures}},\ }\href {https://doi.org/10.1103/PhysRevLett.110.167402}
  {\bibfield  {journal} {\bibinfo  {journal} {Phys. Rev. Lett.}\ }\textbf
  {\bibinfo {volume} {110}},\ \bibinfo {pages} {167402} (\bibinfo {year}
  {2013})}\BibitemShut
  {NoStop}%
\bibitem [{\citenamefont {Gali}(2009)}]{gali_theory_2009}%
  \BibitemOpen
  \bibfield  {author} {\bibinfo {author} {\bibfnamefont {A.}~\bibnamefont
  {Gali}},\ }\bibfield  {title} {\bibinfo {title} {Theory of the neutral
  nitrogen-vacancy center in diamond and its application to the realization of
  a qubit},\ }\href {https://doi.org/10.1103/PhysRevB.79.235210} {\bibfield
  {journal} {\bibinfo  {journal} {Phys. Rev. B}\ }\textbf {\bibinfo {volume}
  {79}},\ \bibinfo {pages} {235210} (\bibinfo {year} {2009})}\BibitemShut {NoStop}%
\bibitem [{\citenamefont {Gardill}\ \emph {et~al.}(2022)\citenamefont
  {Gardill}, \citenamefont {Kemeny}, \citenamefont {Li}, \citenamefont
  {Zahedian}, \citenamefont {Cambria}, \citenamefont {Xu}, \citenamefont
  {Lordi}, \citenamefont {Gali}, \citenamefont {Maze}, \citenamefont {Choy},\
  and\ \citenamefont {Kolkowitz}}]{gardill_super-resolution_2022}%
  \BibitemOpen
  \bibfield  {author} {\bibinfo {author} {\bibfnamefont {A.}~\bibnamefont
  {Gardill}}, \bibinfo {author} {\bibfnamefont {I.}~\bibnamefont {Kemeny}},
  \bibinfo {author} {\bibfnamefont {Y.}~\bibnamefont {Li}}, \bibinfo {author}
  {\bibfnamefont {M.}~\bibnamefont {Zahedian}}, \bibinfo {author}
  {\bibfnamefont {M.~C.}\ \bibnamefont {Cambria}}, \bibinfo {author}
  {\bibfnamefont {X.}~\bibnamefont {Xu}}, \bibinfo {author} {\bibfnamefont
  {V.}~\bibnamefont {Lordi}}, \bibinfo {author} {\bibfnamefont
  {A.}~\bibnamefont {Gali}}, \bibinfo {author} {\bibfnamefont {J.~R.}\
  \bibnamefont {Maze}}, \bibinfo {author} {\bibfnamefont {J.~T.}\ \bibnamefont
  {Choy}},\ and\ \bibinfo {author} {\bibfnamefont {S.}~\bibnamefont
  {Kolkowitz}},\ }\bibfield  {title} {\bibinfo {title} {Super-{Resolution}
  {Airy} {Disk} {Microscopy} of {Individual} {Color} {Centers} in {Diamond}},\
  }\href {https://doi.org/10.1021/acsphotonics.2c00713} {\bibfield  {journal}
  {\bibinfo  {journal} {ACS Photonics}\ }\textbf {\bibinfo {volume} {9}},\
  \bibinfo {pages} {3848} (\bibinfo {year} {2022})}\BibitemShut {NoStop}%
\bibitem [{\citenamefont {Nicolas}\ \emph {et~al.}(2019)\citenamefont
  {Nicolas}, \citenamefont {Delord}, \citenamefont {Huillery},\ and\
  \citenamefont {Hétet}}]{nicolas_optically_2019}%
  \BibitemOpen
  \bibfield  {author} {\bibinfo {author} {\bibfnamefont {L.}~\bibnamefont
  {Nicolas}}, \bibinfo {author} {\bibfnamefont {T.}~\bibnamefont {Delord}},
  \bibinfo {author} {\bibfnamefont {P.}~\bibnamefont {Huillery}},\bibinfo {author} {\bibfnamefont {C.}~\bibnamefont {Pellet-Mary}},\ and\
  \bibinfo {author} {\bibfnamefont {G.}~\bibnamefont {Hétet}},\ }\href {https://doi.org/10.1021/acsphotonics.9b00262}{\bibfield  {journal}
  {\bibinfo  {journal} {ACS Photonics}\ }\textbf {\bibinfo {volume} {6}},\
  \bibinfo {pages} {2413-2420} (\bibinfo {year} {2019})}\BibitemShut {NoStop}%
\bibitem [{\citenamefont {Doherty}\ \emph {et~al.}(2016)\citenamefont
  {Doherty}, \citenamefont {Meriles}, \citenamefont {Alkauskas}, \citenamefont
  {Fedder}, \citenamefont {Sellars},\ and\ \citenamefont
  {Manson}}]{doherty_towards_2016}%
  \BibitemOpen
  \bibfield  {author} {\bibinfo {author} {\bibfnamefont {M.}~\bibnamefont
  {Doherty}}, \bibinfo {author} {\bibfnamefont {C.}~\bibnamefont {Meriles}},
  \bibinfo {author} {\bibfnamefont {A.}~\bibnamefont {Alkauskas}}, \bibinfo
  {author} {\bibfnamefont {H.}~\bibnamefont {Fedder}}, \bibinfo {author}
  {\bibfnamefont {M.}~\bibnamefont {Sellars}},\ and\ \bibinfo {author}
  {\bibfnamefont {N.}~\bibnamefont {Manson}},\ }\bibfield  {title} {\bibinfo
  {title} {Towards a {Room}-{Temperature} {Spin} {Quantum} {Bus} in {Diamond}
  via {Electron} {Photoionization}, {Transport}, and {Capture}},\ }\href
  {https://doi.org/10.1103/PhysRevX.6.041035} {\bibfield  {journal} {\bibinfo
  {journal} {Phys. Rev. X}\ }\textbf {\bibinfo {volume} {6}},\ \bibinfo {pages}
  {041035} (\bibinfo {year} {2016})}\BibitemShut {NoStop}%
\bibitem [{\citenamefont {Rovny}\ \emph {et~al.}(2022)\citenamefont {Rovny},
  \citenamefont {Yuan}, \citenamefont {Fitzpatrick}, \citenamefont {Abdalla},
  \citenamefont {Futamura}, \citenamefont {Fox}, \citenamefont {Cambria},
  \citenamefont {Kolkowitz},\ and\ \citenamefont
  {de~Leon}}]{rovny_nanoscale_2022}%
  \BibitemOpen
  \bibfield  {author} {\bibinfo {author} {\bibfnamefont {J.}~\bibnamefont
  {Rovny}}, \bibinfo {author} {\bibfnamefont {Z.}~\bibnamefont {Yuan}},
  \bibinfo {author} {\bibfnamefont {M.}~\bibnamefont {Fitzpatrick}}, \bibinfo
  {author} {\bibfnamefont {A.~I.}\ \bibnamefont {Abdalla}}, \bibinfo {author}
  {\bibfnamefont {L.}~\bibnamefont {Futamura}}, \bibinfo {author}
  {\bibfnamefont {C.}~\bibnamefont {Fox}}, \bibinfo {author} {\bibfnamefont
  {M.~C.}\ \bibnamefont {Cambria}}, \bibinfo {author} {\bibfnamefont
  {S.}~\bibnamefont {Kolkowitz}},\ and\ \bibinfo {author} {\bibfnamefont
  {N.~P.}\ \bibnamefont {de~Leon}},\ }\bibfield  {title} {\bibinfo {title}
  {Nanoscale covariance magnetometry with diamond quantum sensors},\ }\href
  {https://doi.org/10.1126/science.ade9858} {\bibfield  {journal} {\bibinfo
  {journal} {Science}\ }\textbf {\bibinfo {volume} {378}},\ \bibinfo {pages}
  {1301} (\bibinfo {year} {2022})}\BibitemShut {NoStop}%
\end{thebibliography}
\end{document}